\documentclass[12pt]{article}

\usepackage{a4wide,amssymb,graphicx}

\usepackage{amsmath, amsthm}

\newtheorem{definition}{Definition}
\newtheorem{theorem}{Theorem}
\newtheorem{ass}{Assumption}
\newtheorem{fact}{Fact}
\newtheorem{proposition}{Proposition}
\newtheorem{conj}{Conjecture}
\newtheorem{lemma}{Lemma}

\DeclareMathOperator{\Ai}{Ai}
\DeclareMathOperator{\Bi}{Bi}
\DeclareMathOperator{\Ei}{Ei}
\DeclareMathOperator{\Li}{Li}
\DeclareMathOperator{\erfc}{erfc}
\DeclareMathOperator{\Arg}{Arg}
\DeclareMathOperator{\LerchPhi}{LerchPhi}
\DeclareMathOperator{\e}{e}

\newcommand{\book}[1]{}

\begin{document}

%\centerline{Polygons, polyominoes, and polyhedra}

\title{Limit distributions and scaling functions}\label{Ccr}

\author{\sc Christoph Richard
\\
{\it Fakult\"at f\"ur Mathematik, Universit\"at Bielefeld,}\\
{\it Postfach 10 01 31, 33501 Bielefeld, Germany}}

\maketitle

\begin{abstract}
We discuss the asymptotic behaviour of models of lattice polygons,
mainly on the square lattice. In particular, we focus on limiting area
laws in the uniform perimeter ensemble where, for fixed perimeter,
each polygon of a given area occurs with the same probability.
We relate limit distributions to the scaling behaviour 
of the associated perimeter and area generating functions, 
thereby providing a geometric interpretation of scaling 
functions. To a major extent, this article is a pedagogic 
review of known results.
\end{abstract}

\section{Introduction}

For a given combinatorial class of objects, such as polygons or
polyhedra, the most basic question concerns the number of objects 
of a given size (always assumed to be finite), or an asymptotic 
estimate thereof. Informally stated, in this overview we will 
analyse the refined question:

\begin{center}
What does a typical object look like?
\end{center}

In contrast to the combinatorial question about the number of objects 
of a given size, the latter question is of a probabilistic nature.
For counting parameters in addition to object size, one asks
for their (asymptotic) probability law. To give this question
a meaning, an underlying ensemble has to be specified. The simplest choice
is the uniform ensemble, where each object of a given size occurs with equal
probability. 

For self-avoiding polygons on the square lattice, size may be the number 
of edges of the polygon, and 
an additional counting parameter may be the area enclosed by the polygon. 
We will call this ensemble the \emph{fixed perimeter ensemble}\index{fixed 
perimeter ensemble}. For the
\emph{uniform} fixed perimeter ensemble, one assumes that, for a fixed number 
of edges, each polygon occurs with the same probability. Another 
ensemble, which we will call the \emph{fixed area ensemble}\index{fixed area ensemble}, 
is obtained with size 
being the polygon area, and the number of edges being an additional 
counting parameter. For the \emph{uniform} fixed area ensemble, one assumes that, 
for fixed area, each polygon occurs with the same probability. 

To be specific, let $p_{m,n}$ denote the number of square lattice 
self-avoiding polygons of half-perimeter $m$ and area $n$. Discrete 
random variables $\widetilde X_m$ of area in the uniform fixed 
perimeter ensemble and of perimeter $\widetilde Y_n$ in the uniform 
fixed area ensemble are defined by
\begin{displaymath}
\mathbb P(\widetilde X_m=n) =\frac{p_{m,n}}{\sum_n p_{m,n}},\qquad
\mathbb P(\widetilde Y_n=m) =\frac{p_{m,n}}{\sum_m p_{m,n}}.
\end{displaymath}
We are interested in an asymptotic description of these probability
laws, in the limit of infinite object size.

\medskip

In statistical physics, certain non-uniform ensembles are important. 
For fixed object size, the probability of an object with value $n$
of the counting parameter (such as the area of a polygon) 
may be proportional to $a^n$, for some non-negative parameter 
$a=\e^{-\beta E}$ of non-uniformity. Here $E$
is the energy of the object, and $\beta=1/(k_B T)$, where $T$ is the
temperature, and $k_B$ denotes Boltzmann's constant. A qualitative 
change in the 
behaviour of typical objects may then be reflected in a qualitative 
change in the probability law of the counting parameter w.r.t.~$a$. Such a 
change is an indication of a phase transition, i.e., a non-analyticity
in the free energy of the corresponding ensemble.

For self-avoiding polygons in the fixed perimeter ensemble, let $q$ denote 
the parameter of non-uniformity,
\begin{displaymath}
\mathbb P(\widetilde X_m(q)=n) =\frac{p_{m,n}q^n}{\sum_n p_{m,n}q^n}.
\end{displaymath}
Polygons of large area are suppressed in probability for small values 
of $q$, such that one expects a typical self-avoiding polygon to 
closely resemble a branched polymer. Likewise, for large values of 
$q$, a typical polygon is expected to be inflated, closely resembling 
a ball (or square) shape. Let us define the \emph{ball-shaped phase} 
by the condition that the mean area of a polygon grows quadratically 
with its perimeter. The ball-shaped phase occurs for $q>1$ \cite{cr:FGW91}. 
Linear growth of the mean area w.r.t. perimeter is expected to occur 
for all values $0<q<1$. This phase called the \emph{branched polymer 
phase}. Of particular interest is the point $q=1$, at which a phase 
transition occurs \cite{cr:FGW91}. This transition is called a 
\emph{collapse transition}\index{collapse transition}. Similar 
considerations apply for self-avoiding polygons in the fixed area ensemble, 
\begin{displaymath}
\mathbb P(\widetilde Y_n(x)=m) =\frac{p_{m,n}x^m}{\sum_m p_{m,n}x^m},
\end{displaymath}
with parameter of non-uniformity $x$, where $0<x<\infty$.

For a given model, these effects may be studied using data from exact 
or Monte-Carlo enumeration and series extrapolation techniques. Sometimes, the 
underlying model is exactly solvable, i.e., it obeys 
a combinatorial decomposition, which leads to a recursion for 
the counting parameter. In that case, its (asymptotic) behaviour 
may be extracted from the recurrence.

A convenient tool is generating functions.
The combinatorial information about the number of objects of a given 
size is coded in a one-variable (ordinary) generating function, typically
of positive and finite radius of convergence. Given the generating 
function of the counting problem, the asymptotic behaviour
of its coefficients can be inferred from the leading singular behaviour
of the generating function. This is determined by the
location and nature of the singularity of the generating function closest
to the origin. There are elaborate techniques for studying this 
behaviour exactly \cite{cr:FS06} or numerically \cite{cr:G89}\book{, see chapter 8}.

The case of additional counting parameters leads to a multivariate generating 
function. For self-avoiding polygons, the half-perimeter and area generating
function is 
\begin{displaymath}
P(x,q)=\sum_{m,n}p_{m,n}x^mq^n.
\end{displaymath}
For a fixed value of a non-uniformity parameter $q_0$, where 
$0<q_0\le1$, let $x_0$ be the radius of convergence of $P(x,q_0)$. 
The asymptotic law of the counting parameter is encoded in the 
singular behaviour of the generating function $P(x,q)$ 
about $(x_0,q_0)$. If locally about $(x_0,q_0)$ the nature 
of the singularity of $P(x,q)$ does not change, then 
distributions are expected to be concentrated, with a Gaussian limit law.
This corresponds to the physical intuition that fluctuations 
of macroscopic quantities are asymptotically negligible 
away from phase transition points. If the nature of the 
singularity does change locally, we expect non-concentrated 
distributions, resulting in non-Gaussian limit laws. This
is expected to be the case at phase transition points.

\begin{figure}[htb]
\begin{center}
\begin{minipage}[b]{0.95\textwidth}
\center{\includegraphics[width=4cm]{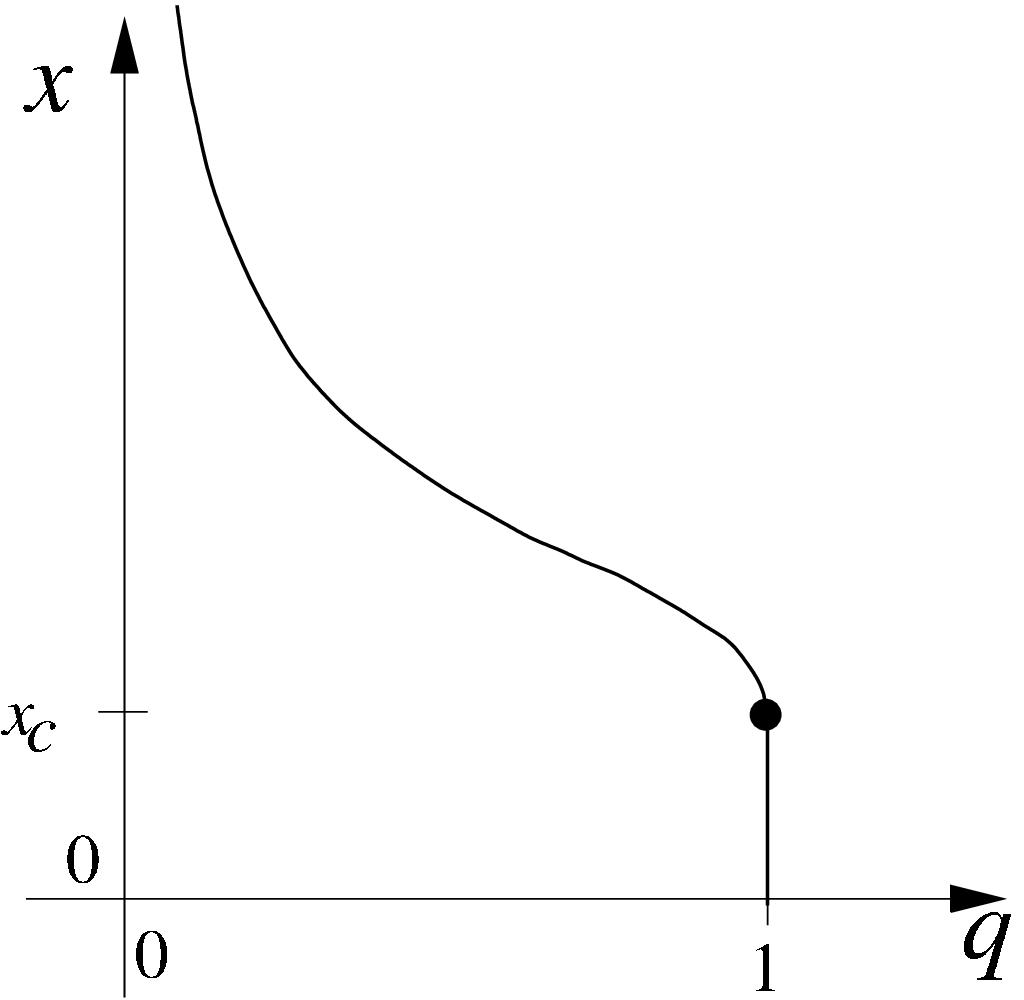}}
\end{minipage}
\end{center}
\caption{\label{Fcr:stair}
\small Singularity diagram of a typical polygon model counted by 
half-perimeter and area, with $x$ conjugate to half-perimeter and 
$q$ conjugate to area.}
\end{figure}

Qualitative information about the singularity structure is 
given by the \emph{singularity diagram}\index{singularity diagram} (also called the 
\emph{phase diagram}\index{phase diagram})\book{, compare chapter 2}.
It displays the region of convergence of the two-variable generating 
function, i.e., the set of points $(x,q)$ in the closed upper right quadrant
of the plane, such that the generating function $P(x,q)$ converges. 
The set of boundary points with positive coordinates is a set of singular 
points of $P(x,q)$, called the \emph{critical curve}. See Figure~\ref{Fcr:stair} 
for a sketch of the singularity diagram of a typical polygon model such 
as self-avoiding polygons, counted by half-perimeter and area, with 
generating function $P(x,q)$ as above.
There appear two lines of singularities, which intersect at the 
point $(x,q)=(x_c,1)$. Here $x_c$ is the radius of 
convergence of the half-perimeter generating function $P(x,1)$, 
also called the \emph{critical point}. The nature of a singularity 
does not change along each of the two lines, and the intersection 
point $(x,q)=(x_c,1)$ of the two lines is a phase transition point. 
For $0<q<1$ fixed, denote by $x_c(q)$ the radius of convergence 
of $P(x,q)$. The branched polymer phase for the fixed perimeter ensemble 
$0<q<1$ (and also for the corresponding fixed area ensemble) is 
asymptotically described by the singularity of $P(x,q)$ about $(x_c(q),q)$.
In the ball-shaped phase $q>1$ of the fixed perimeter ensemble, the 
(ordinary) generating function does not seem the right object to study,
since it has zero radius of convergence for fixed $q>1$.
The singularity of $P(x,q)$ about $(x,1)$ describes, for $0<x<x_c$,
a ball-shaped phase in the fixed area ensemble, with a \emph{finite}
average size of a ball.

For points $(x,q)$ within the region of convergence, both $x$ and $y$ 
positive, the generating function $P(x,q)$ is finite and positive. Thus, 
such points may be interpreted as parameters in a mixed infinite ensemble
\begin{displaymath}
\mathbb P(\widetilde X(x,q)=(m,n)) =\frac{p_{m,n}x^mq^n}{\sum_{m,n} 
p_{m,n}x^mq^n}.
\end{displaymath}

The limiting law of the counting parameter in the fixed area
or fixed perimeter ensemble can be extracted from the leading 
singular behaviour of the two-variable generating function.
There are two different approaches to the problem. The first one
consists in analysing, for fixed non-uniformity parameter $a$, 
the singular behaviour of the remaining one-parameter generating 
function and its derivatives w.r.t.~$a$. This method is also called 
the \emph{method of moments}. It can be successfully applied
in the fixed perimeter ensemble at 
the phase transition point. Typically, this results in 
non-concentrated distributions.

The second approach derives an asymptotic approximation of the two-variable 
generating function. Away from a phase transition point, such an 
approximation can be obtained for some classes of models, typically resulting 
in concentrated distributions, with a Gaussian law for the centred 
and normalised random variable. However, it is usually difficult to 
extract such information at a phase transition point. The theory of 
tricritical scaling seeks to fill this gap, by suggesting and justifying 
a particular ansatz for an approximation using scaling functions. 
Knowledge of the approximation may imply knowledge of the quantities 
analysed in the first approach.

In the following, we give an overview of these two
approaches. For the first approach, summarised by the title \emph{limit
distributions}\index{limit distribution}, there are a number of rigorous 
results, which we will discuss. The second approach, summarised by the title 
\emph{scaling functions}\index{scaling function}, is
less developed. For that reason, our
presentation will be more descriptive, stating important open questions.
We will stress connections between the two approaches, thereby
providing a probabilistic interpretation of scaling functions in terms
of limit distributions. 

\section{Polygon models and generating functions}

Models of polygons, polyominoes or polyhedra have been studied 
intensively on the square and cubic lattices. It is believed that the 
leading asymptotic behaviour of such models, such as the type
of limit distribution or critical exponents, is independent of 
the underlying lattice.

In two dimensions, a number of models of square lattice polygons have
been enumerated according to perimeter and area and other parameters, 
see \book{chapter 3 and} \cite{cr:B96} for a review of models with an 
exact solution. The majority of such models has an algebraic perimeter
generating function. We mention prudent polygons \cite{cr:S08,cr:D08,cr:BM08} 
as a notable exception. Of particular importance for polygon models is the fixed 
perimeter ensemble, since it models
two-dimensional vesicle collapse. Another important ensemble is the fixed
area ensemble, which serves as a model of ring polymers. The fixed area 
ensemble may also describe percolation and cluster growth. For example, 
staircase polygons are models of \emph{directed compact percolation} 
\cite{cr:E89,cr:ET90,cr:ET94,cr:EG95,cr:BE99,cr:K03}. This may be compared to the exactly 
solvable case of percolation on a tree \cite{cr:Grimmett99}.
The model of self-avoiding polygons is conjectured to describe the hull of 
critical percolation clusters \cite{cr:LSW02}.

In addition to perimeter, other counting parameters have been studied, such as
width and height, generalisations of area \cite{cr:Rich05}, 
radius of gyration \cite{cr:J05,cr:L06}, number of nearest-neighbour interactions 
\cite{cr:BEGGLOW98}, last column height \cite{cr:B96}, and site perimeter 
\cite{cr:DGV87,cr:BR03}. Also, motivated by applications in chemistry, symmetry 
subclasses of polygon models have been analysed \cite{cr:LER98,cr:LR01,cr:GL05, 
cr:RST07}. Whereas this gives rise to a number of different ensembles, 
only a few of them have been asymptotically studied. Not all of them 
display phase transitions.

In three dimensions, models of polyhedra on the cubic lattice 
have been enumerated according to perimeter, surface area and 
volume, see \cite{cr:MaSl93,cr:V98,cr:AB06} and the discussion in section
\ref{cr:sec3d}. Various ensembles may be defined, 
such as the fixed surface area ensemble and the fixed volume 
ensemble. The fixed surface area ensemble serves as a model 
of three-dimensional vesicle collapse \cite{cr:W93}. 

In this chapter, we will consider models of square lattice
polygons, counted by half-perimeter and area. Let $p_{m,n}$ 
denote the (finite) number of such polygons of half-perimeter $m$ 
and area $n$. The numbers $p_{m,n}$ will always
satisfy the following assumption.

\begin{ass}\label{Acr:pol}
For $m,n\in\mathbb N_0$, let non-negative integers $p_{m,n}\in\mathbb N_0$ be given.
The numbers $p_{m,n}$ are assumed to satisfy the following properties.
\begin{itemize}
\item[{\it i)}] There exist positive constants $A,B>0$ such that 
$p_{m,n}=0$ if $n\le Am$ or if $n\ge Bm^2$.
\item[{\it ii)}] The sequence $\left(\sum_{n}p_{m,n}
\right)_{m\in\mathbb N_0}$ has infinitely many positive
elements and grows at most exponentially.
\end{itemize}
\end{ass}

\noindent \textbf{Remarks.} \textit{i)} A sequence $(a_n)_{n\in
\mathbb N_0}$ is said to \emph{grow at most exponentially}, if 
there are positive constants $C$, $\mu$ such that $|a_n|\le C 
\mu^n$ for all $n$.\\
\textit{ii)} Condition $i)$ reflects the geometric constraint that the 
area of a polygon grows at most quadratically and at least linearly 
with its perimeter. For self-avoiding polygons, we have $n\ge m-1$. 
Since $p_{m,n}=0$ if $m<2$, we may choose $A=1/3$. Since 
$n\le m^2/4$ for self-avoiding polygons, we may choose $B=1/3$. 
Condition \textit{ii)} is a natural condition on the growth of the 
number of polygons of a given perimeter. For self-avoiding polygons, we
may choose $C=1$ and $\mu=16$.\\
\textit{iii)} For models with counting parameters different from area, or for
models in higher dimensions, a modified assumption holds, with the growth
condition \textit{i)} being replaced by $n\le A m^{k_0}$ and $n\ge B m^{k_1}$, for
appropriate values of $k_0$ and $k_1$. Counting parameters statisfying 
$p_{m,n}=0$ for $n\ge B m^k$ are called \emph{rank $k$ parameters} 
\cite{cr:Duchon99}.

\medskip

The above assumption imposes restrictions on the generating
function of the numbers $p_{m,n}$. These explain the qualitative 
form of the singularity diagram Figure~\ref{Fcr:stair}.

\begin{proposition}\label{cr:prop:pol}
For numbers $p_{m,n}$, let Assumption \ref{Acr:pol} be satisfied. Then,
the generating function $P(x,q)=\sum_{m,n}p_{m,n}x^m q^n$ has the
following properties.
\begin{itemize}
\item[\textit{i)}]
The generating function $P(x,q)$ satisfies for $k\in\mathbb N$
\begin{equation*}
A^k\left(x\frac{\partial}{\partial x}\right)^{k}P(x,q)\ll
\left(q\frac{\partial}{\partial q}\right)^kP(x,q)\ll 
B^k \left(x\frac{\partial}{\partial x}\right)^{2k}P(x,q),
\end{equation*}
where $\ll$ denotes coefficient-wise domination.

\item[\textit{ii)}]
The evaluation $P(x,1)$ is a power series with radius of
convergence $x_c$, where $0<x_c\le1$. 

\item[\textit{iii)}]
The generating function $P(x,q)$ diverges, if $x\ne0$ and $|q|>1$. 
It converges, if $|q|<1$ and $|x|<x_cq^{-A}$. In particular, for 
$k\in\mathbb N_0$, the evaluations
\begin{equation*}
\left.\frac{\partial^k}{\partial x^k}P(x,q)\right|_{x=x_c}
\end{equation*} 
are power series with radius of convergence 1.

\item[\textit{iv)}]
For $k\in\mathbb N_0$, the evaluations
\begin{equation*}
\left.\frac{\partial^k}{\partial q^k}P(x,q)\right|_{q=1}
\end{equation*} 
are power series with radius of convergence $x_c$. They
satisfy, for $|x|<x_c$,
\begin{displaymath}
\left.\frac{\partial^k}{\partial q^k}P(x,q)\right|_{q=1} = 
\lim_{\substack{q\to1\\-1<q<1}}\frac{\partial^k}{\partial q^k}P(x,q).
\end{displaymath}

\end{itemize}

\end{proposition}

\begin{proof}[sketch]
The domination formula follows immediately from condition \textit{i)}. 
The existence of the evaluations at $q=1$ and $x=x_c$ as formal 
power series also follows from condition \textit{i)}. Condition \textit{ii)}
ensures that $0<x_c\le1$ for the radius of convergence of $P(x,1)$.
Equality of the radii of convergence for the derivatives follows from 
condition \textit{i)} by elementary estimates. The claimed analytic 
properties of $P(x,q)$ follow from conditions \textit{i)} and \textit{ii)} by 
elementary estimates. The claimed left-continuity of the derivatives in 
\textit{iv)} is implied by Abel's continuity theorem for real power series.
\qed
\end{proof}

\noindent {\bf Remarks.} {\it i)} Proposition \ref{cr:prop:pol} implies 
that the critical curve $x_c(q)$ satisfies for $0<q<1$ the estimate 
$x_c(q)\ge x_cq^{-A}$. For self-avoiding polygons, the critical curve 
$x_c(q)$ is continuous for $0<q<1$. This follows from a certain 
supermultiplicative inequality for the numbers $p_{m,n}$ by 
convexity arguments \cite{cr:Janse05}.\\
{\it ii)} Of central importance in the sequel will be the power series
\begin{equation}\label{Ecr:gengk}
g_k(x)=\frac{1}{k!}\left.\frac{\partial^k}
{\partial q^k}P(x,q)\right|_{q=1}. 
\end{equation} 
They are called \emph{factorial moment generating functions}\index{moment
generating function}, for reasons which will become clear later. 

\medskip

We continue studying analytic properties of the factorial
moment generating functions. In the following, the notation 
$x\nearrow x_0$ denotes the limit $x\to x_0$ for sequences 
$(x_n)$ satisfying $|x_n|<x_0$. The notation 
$f(x)\sim g(x)$ as $x\nearrow x_0$ means that $g(x)\ne0$ in
a left neigbourhood of $x_0$ and that
$\lim_{x\nearrow x_0}f(x)/g(x)=1$. Likewise, $a_m\sim b_m$
as $m\to\infty$ for sequences $(a_m),(b_m)$ means that $b_m\ne0$
for almost all $m$ and $\lim_{m\to\infty} a_m/b_m=1$. 
The following lemma is a standard result.

\begin{lemma}\label{cr:prop:asfinmom}
Let $(a_m)_{m\in\mathbb N_0}$ be a sequence of real
numbers, which asymptotically satisfy
\begin{equation}\label{Ecr:gencfasym}
a_m\sim A x_c^{-m} 
m^{\gamma-1}\qquad (m\to\infty),
\end{equation}
for real numbers $A,x_c,\gamma$, where $A\ne0$ and $x_c>0$. 

Then, the generating function $g(x)=\sum_{m=0}^\infty a_m x^m$
has radius of convergence $x_c$.
If $\gamma\notin\{0,-1,-2,\ldots\}$, then there exists a power series
$g^{(reg)}(x)$ with radius of convergence strictly larger than 
$x_c$, such that $g(x)$ satisfies
\begin{equation}\label{Ecr:gengfasym}
\left(g(x)-g^{(reg)}(x)\right)\sim \frac{A\,\Gamma(\gamma)}{(1-x/x_c)^{\gamma}}
\qquad (x\nearrow x_c),
\end{equation}
where $\Gamma(z)$ denotes the Gamma function.
\qed
\end{lemma}

\noindent {\bf Remarks.} {\it i)} The above lemma can
be proved using the analytic properties of the polylog function \cite{cr:F99}. 
If $\gamma\in\{0,-1,-2,\ldots\}$, an asymptotic form
similar to Eq.~(\ref{Ecr:gengfasym}) is valid, which involves logarithms.\\
{\it ii)} The function $g^{(reg)}(x)$ in the above lemma 
is not unique. For example, if $\gamma>0$, any polynomial in 
$x$ may be chosen. We demand $g^{(reg)}(x)\equiv0$ in that case. 
If $\gamma<0$ and $g^{(reg)}(x)$ is restricted to be a polynomial, 
it is uniquely defined. If $-1<\gamma<0$, we have $g^{(reg)}(x)\equiv g(x_c)$. 
In the general case, the polynomial has degree $\lfloor -\gamma 
\rfloor$, compare \cite{cr:F99}. In the following, we will demand uniqueness
by the above choice. The power series $g^{(sing)}(x):=\left(g(x)-g^{(reg)}(x)\right)$
is then called \emph{the singular part} of $g(x)$.

\medskip

Conversely, let a power series $g(x)$ with radius of convergence 
$x_c$ be given. In order to conclude from Eq.~(\ref{Ecr:gengfasym})
the behaviour Eq.~(\ref{Ecr:gencfasym}), certain additional 
analyticity assumptions on $g(x)$ have to be satisfied. To this 
end, a function $g(x)$ is called \emph{$\Delta(x_c,\eta,\phi)$-regular} 
(or simply \emph{$\Delta$-regular)} \cite{cr:FFK05}, if there is a 
positive real number $x_c>0$, such that $g(x)$ is analytic in the 
\emph{indented disc} $\Delta(x_c,\eta,\phi):=\{z\in\mathbb C:|z|\le x_c+\eta, 
|\Arg(z-x_c)|\ge\phi\}$, for some $\eta>0$ and some $\phi$, where 
$0<\phi<\pi/2$. Note that $x_c\notin\Delta$, where we adopt the 
convention $\Arg(0)=0$. The point $x=x_c$ is the only point for $|x|\le x_c$, 
where $g(x)$ may possess a singularity.

\begin{lemma}[\cite{cr:FO90}]\label{Ecr:transfergx}
Let the function $g(x)$ be $\Delta$-regular and assume that
\begin{displaymath}
g(x)\sim \frac{1}{\left(1-x/x_c\right)^\gamma} \qquad (x\to x_c 
\,\text{ in }\, \Delta).
\end{displaymath}
If $\gamma\notin\{0,-1,-2,\ldots\}$, we then have
\begin{displaymath}
[x^m]g(x)\sim \frac{1}{\Gamma(\gamma)} x_c^{-m}m^{\gamma-1} 
\qquad(m\to\infty),
\end{displaymath}
where $[x^m]g(x)$ denotes the Taylor coefficient of $g(x)$ 
of order $m$ about $x=0$.
\qed
\end{lemma}

\noindent \textbf{Remarks.} \textit{i)} Note that the 
coefficients of the function $f(x)=(1-x/x_c)^{-\gamma}$ 
with real exponent $\gamma\notin\{0,-1,-2,\ldots\}$ satisfy
\begin{equation}\label{Ecr:trans}
[x^m]f(x)\sim\frac{1}{\Gamma(\gamma)}x_c^{-m}
m^{\gamma-1}\qquad (m\to\infty).
\end{equation}
This may be seen by an application of the binomial series and 
Stirling's formula. For functions $g(x)\sim f(x)$, the assumption 
of $\Delta$-regularity for $g(x)$ ensures that the 
same asymptotic estimate holds for the coefficients of $g(x)$. \\
\textit{ii)} Theorems of the above type are called 
\emph{transfer theorems}\index{transfer theorem} \cite{cr:FO90, cr:FS06}.
The set of $\Delta$-regular functions with singularities of the 
above form is closed under addition, multiplication, differentiation, 
and integration \cite{cr:FFK05}. \\
\textit{iii)} The case of a finite number of singularities on the
circle of convergence can be treated by a straightforward extension
of the above result \cite{cr:FO90, cr:FS06}.

\medskip

Lemma \ref{cr:prop:asfinmom} implies a particular singular behaviour
of the factorial moment generating functions, if the numbers
$p_{m,n}$ satisfy certain typical asymptotic estimates. We
write $(a)_k=a\cdot(a-1)\cdot\ldots\cdot(a-k+1)$ to denote
the lower factorial.

\begin{proposition}\label{cr:prop:gksing}
For $m,n\in\mathbb N_0$, let real numbers $p_{m,n}$ be given.
Assume that the numbers $p_{m,n}$ asymptotically satisfy, 
for $k\in\mathbb N_0$,
\begin{equation}\label{Ecr:ampli}
\frac{1}{k!}\sum_n (n)_k p_{m,n}\sim A_k x_c^{-m} 
m^{\gamma_k-1}\qquad (m\to\infty),
\end{equation}
for real numbers $A_k,x_c,\gamma_k$, where $A_k>0$, $x_c>0$, and
$\gamma_k\notin \{0,-1,-2,\ldots\}$.

Then, the factorial moment generating functions 
$g_k(x)$ satisfy
\begin{equation}
g_k^{(sing)}(x)\sim \frac{f_k}{(1-x/x_c)^{\gamma_k}}\qquad (x\nearrow x_c),
\end{equation}
where $f_k=A_k\,\Gamma(\gamma_k)$. \qed
\end{proposition}

\noindent \textbf{Remarks.} \textit{i)} The above assumption
on the growth of the coefficients in Eq.~(\ref{Ecr:ampli})
is typical for polygon models, with $\gamma_k=(k-\theta)/\phi$, 
and $\phi>0$.\\
\textit{ii)} If the numbers $p_{m,n}$ satisfy, in addition to Eq.÷(\ref{Ecr:ampli}), 
Condition $i)$ of Assumption \ref{Acr:pol}, this implies for
exponents of the form $\gamma_k=(k-\theta)/\phi$, where $\phi>0$,
the estimate $1/2\le\phi\le 1$. \\
\textit{iii)} The proposition implies that the singular
part of the factorial moment generating function $g_k(x)$ is 
asymptotically equal to the singular part of the corresponding 
(ordinary) moment generating function,
\begin{equation*}
\left(\left.\frac{\partial^k}
{\partial q^k}P(x,q)\right|_{q=1}\right)^{(sing)} 
\sim 
\left(\left.\left(q\frac{\partial}
{\partial q}\right)^kP(x,q)\right|_{q=1}\right)^{(sing)} 
\qquad (x\nearrow x_c).
\end{equation*}

\medskip

We give a list of exponents and area limit distributions for a 
number of polygon models. An asterisk denotes that corresponding results
rely on a numerical analysis. It appears that the value
$(\theta,\phi)=(1/3,2/3)$ arises for a large number of models. 
Furthermore, the exponent $\gamma_0$ seems to determine the area
limit law. These two observations will be explained
in the following section.

\begin{table}
\begin{center}
\begin{tabular}{|c|c|c|c|c|}
\hline
{\it Model} & $\phi$ & $\theta$ & $\gamma_0$ & {\it Area limit law} \\\hline \hline
rectangles&&&&\\
convex polygons & \raisebox{1.5ex}[-1.5ex]{$\frac{1}{2}$} & \raisebox{1.5ex}[-1.5ex]{$-1$}
& \raisebox{1.5ex}[-1.5ex]{$2$}& \raisebox{1.5ex}[-1.5ex]{$\beta_{1,1/2}$} \\\hline
Ferrers diagrams&&&&\\
stacks& \raisebox{1.5ex}[-1.5ex]{$\frac{1}{2}$} & \raisebox{1.5ex}[-1.5ex]{$-\frac{1}{2}$}
& \raisebox{1.5ex}[-1.5ex]{$1$}& \raisebox{1.5ex}[-1.5ex]{Gaussian} \\\hline
staircase polygons&&&&\\
bargraph polygons&&&&\\
column-convex polygons&&&&\\
directed column-convex polygons& \raisebox{1.5ex}[-1.5ex]{$\frac{2}{3}$} & \raisebox{1.5ex}[-1.5ex]{$\frac{1}{3}$}
& \raisebox{1.5ex}[-1.5ex]{$-\frac{1}{2}$}& \raisebox{1.5ex}[-1.5ex]{Airy}\\
diagonally convex directed polygons&&&&\\
rooted self-avoiding polygons$^{*}$ &&&&\\\hline
directed convex polygons & $\frac{2}{3}$& $-\frac{1}{3}$&$\frac{1}{2}$& meander\\\hline
diagonally convex polygons$^{*}$ &  &  & $-\frac{1}{2}$& \\\hline
three-choice polygons &  &  & 0& \\\hline
\end{tabular}
\end{center}
\caption{
\small Exponents and area limit laws for prominent polygon models. 
An asterisk denotes a numerical analysis.}
\end{table}

\section{Limit distributions}\index{limit distribution}

In this section, we will concentrate on models of square lattice polygons
in the fixed perimeter ensemble, and analyse their area law.
The uniform ensemble is of particular interest, since non-Gaussian
limit laws usually appear, due to expected phase transitions at $q=1$.
For non-uniform ensembles $q\ne1$, Gaussian limit laws are expected, due to the
absence of phase transitions. 

There are effective techniques for the 
uniform ensemble, since the relevant generating functions are 
typically algebraic. This is different from the fixed area ensemble, 
where singularities are more difficult to analyse. It will 
turn out that the dominant singularity of the perimeter generating 
function determines the limiting area law of the model. We will 
first discuss several examples with different type of singularity. Then,
we will describe a general result, by analysing classes of $q$-difference
equations (see e.g. \cite{cr:DRSZ03}), which exactly solvable polygon models obey. 
Whereas in the case $q\ne1$ their theory is developed to some extent, the 
case $q=1$ is more difficult to analyse. Motivated by the typical 
behaviour of polygon models, we assume that a $q$-difference equation 
reduces to an algebraic equation as $q$ approaches unity, and then 
analyse the behaviour of its solution about $q=1$.

Useful background concerning a probabilistic analysis of
counting parameters of combinatorial structures can be found in
\cite[Ch~IX]{cr:FS06}. See \cite[Ch~1]{cr:O74} and \cite[Ch~1]{cr:BH86} for 
background about asymptotic expansions. For properties of 
formal power series, see \cite[Ch~1.1]{cr:GJ83}. A useful reference
on the Laplace transform, which will appear below, is \cite{cr:D74}.

\subsection{An illustrative example: Rectangles}\index{rectangles}

\subsubsection{Limit law of area}

Let $p_{m,n}$ denote the number of rectangles of half-perimeter $m$ 
and area $n$. Consider the uniform fixed perimeter ensemble, with
a discrete random variable of area $\widetilde X_m$ defined by
\begin{equation}\label{Ecr:arearv}
\mathbb P(\widetilde X_m=n)=\frac{p_{m,n}}{\sum_np_{m,n}}.
\end{equation}
The $k$-th moments of $\widetilde X_m$ are given explicitly by
\begin{equation*}
\begin{split}
\mathbb E [\widetilde X_m^k] &= \sum_{l=1}^{m-1} (l(m-l))^k \frac{1}{m-1}\\ 
&\sim m^{2k} \int_0^1 (x(1-x))^k {\rm d}x
=\frac{(k!)^2}{(2k+1)!}m^{2k} \qquad (m\to\infty),
\end{split}
\end{equation*}
where we approximated the Riemann sum by an integral, using the
Euler-MacLaurin summation formula.
Thus, the random variable $\widetilde X_m$ has mean $\mu_m\sim m^2/6$ and variance
$\sigma_m^2\sim m^4/180$. Since the sequence of random 
variables $(\widetilde X_m)$ does not satisfy the concentration property 
$\lim_{m\to\infty}\sigma_m/\mu_m=0$, we expect a non-trivial limiting 
distribution. Consider the normalised random variable 
\begin{equation}\label{Ecr:recxm}
X_m=\frac{2}{3}\frac{\widetilde X_m}{\mu_m}=4\frac{\widetilde X_m}{m^2}.
\end{equation}
Since the moments of $X_m$ converge as $m\to\infty$, and
the limit sequence $M_k:=\lim_{m_\to\infty} \mathbb E[X_m^k]$
satisfies the Carleman condition $\sum_k(M_{2k})^{-1/(2k)}=\infty$, 
they define \cite[Ch~4.5]{cr:Ch74} a unique random variable 
$X$ with moments $M_k$. Its moment generating function 
$M(t)=\mathbb E[\e^{-tX}]$ is readily obtained as
\begin{equation*}
M(t)=\sum_{k=0}^\infty \frac{\mathbb E [X^k]}{k!}(-t)^k = 
\frac{1}{2}\sqrt{\frac{\pi}{t}} \e^{t}\mbox{erf} \left( \sqrt{t}\right).
\end{equation*}
The corresponding probability distribution $p(x)$ is obtained by
an inverse Laplace transform, and is given by
\begin{equation}\label{Ecr:beta}
p(x)=\left\{ \begin{array}{cr} \frac{1}{2\sqrt{1-x}} & 0\le x\le 1 \\ 
0 & x>1 \end{array}\right..
\end{equation} 
This distribution is known as the beta distribution $\beta_{1,1/2}$.
Together with \cite[Thm~4.5.5]{cr:Ch74}, we arrive at the following 
result.

\begin{theorem}
The area random variable $\widetilde X_m$ of rectangles 
Eq.~(\ref{Ecr:arearv}) has mean $\mu_m\sim m^2/6$ and 
variance $\sigma_m^2\sim m^4/180$. The normalised
random variables $X_m$ Eq.~(\ref{Ecr:recxm}) converge in 
distribution to a continuous random variable with limit 
law $\beta_{1,1/2}$ Eq.~(\ref{Ecr:beta}). We also have moment
convergence.
\qed
\end{theorem}

\subsubsection{Limit law via generating functions}

We now extract the limit distribution using generating functions. 
Whereas the derivation is less direct than the previous 
approach, the method applies to a number of other cases, 
where a direct approach fails. Consider the half-perimeter 
and area generating function $P(x,q)$ for rectangles,
\begin{equation*}
P(x,q)=\sum_{m,n} p_{m,n}x^mq^n.
\end{equation*}
The factorial moments of the area random variable $\widetilde X_m$
Eq.~(\ref{Ecr:arearv}) are obtained from the generating function via
\begin{equation*}
\mathbb E[(\widetilde X_m)_k] = \frac{\sum_n (n)_kp_{m,n}}{\sum_n p_{m,n}} 
= \frac{[x^m]\left.\frac{\partial^k}{\partial q^k}
P(x,q)\right|_{q=1}}{[x^m]P(x,1)},
\end{equation*}
where $(a)_k=a\cdot(a-1)\cdot\ldots\cdot(a-k+1)$ is the lower
factorial. The generating function $P(x,q)$ satisfies 
\cite[Eq.~5.1]{cr:Rich02} the linear $q$-difference equation \cite{cr:DRSZ03}
\begin{equation}\label{Ecr:recfeq}
P(x,q)= x^2 q P(qx,q)+\frac{x^2q(1+qx)}{1-qx}.
\end{equation}
Due to the particular structure of the functional equation,
the area moment generating functions
\begin{equation*}
g_k(x)=\frac{1}{k!}\left.\frac{\partial^k}
{\partial q^k}P(x,q)\right|_{q=1}
\end{equation*} 
are rational functions and can be computed recursively 
from the functional equation, by repeated 
differentiation w.r.t.~$q$ and then setting $q$=1. 
(Such calculations are easily performed with a 
computer algebra system.) This gives, in particular,
\begin{equation*}
\begin{split}
g_0(x) &= \frac{x^2}{(1-x)^2}, \qquad g_1(x) = \frac{x^2}{(1-x)^4},\\
g_2(x) &= \frac{2x^3}{(1-x)^6}, \qquad g_3(x) = \frac{6x^4}{(1-x)^8},\\
g_4(x) &= \frac{x^4(1+22x+x^2)}{(1-x)^{10}},\qquad 
g_5(x) = \frac{12x^5(1+8x+x^2)}{(1-x)^{12}}.
\end{split}
\end{equation*}
Whereas the exact expressions get messy for increasing $k$, their
asymptotic form about their singularity $x_c=1$ is
simply given by
\begin{equation}\label{Ecr:recsing}
g_k(x)\sim\frac{k!}{(1-x)^{2k+2}} \qquad (x\to1).
\end{equation}
The above result can be inferred from the functional equation, 
which induces a recursion for the functions $g_k(x)$, 
which in turn can be asymptotically analysed. This method 
is called \emph{moment pumping} \cite{cr:FPV98}. Below, we 
will extract the above asymptotic behaviour by the method 
of dominant balance.

The asymptotic behaviour of the
moments of $\widetilde X_m$ can be obtained from singularity analysis
of generating functions, as described in Lemma \ref{Ecr:transfergx}. 
Using the functional equation, it can be shown that all functions 
$g_k(x)$ are Laurent series about $x=1$, with a finite number of terms. 
Hence the remark following Lemma \ref{Ecr:transfergx} implies for the 
(factorial) moments of the random variable $X_m$ Eq.~(\ref{Ecr:recxm}) 
the expression
\begin{equation*}
\frac{\mathbb E[(X_m)^k]}{k!}\sim \frac{\mathbb E[(X_m)_k]}{k!}
\sim\frac{k!}{\Gamma(2k+2)}=\frac{k!}{(2k+1)!} \qquad (m\to\infty),
\end{equation*}
in accordance with the previous derivation. 

On the level of the moment generating function, an application of Watson's 
lemma \cite[Sec~4.1]{cr:BH86} shows that the coefficients $k!$ in 
Eq.~(\ref{Ecr:recsing}) appear in the asymptotic expansion of a certain 
Laplace transform of the (entire) moment generating function $\mathbb E[\e^{-tX}]$,
\begin{equation*}
\int_0^\infty \e^{-st} \left(\sum_{k\ge0}\frac{\mathbb E[X^k]}{k!} (-t^2)^{k}\right)
t \,{\rm d}t
 \sim \sum_{k\ge0} (-1)^k k! s^{-(2k+2)} \qquad (s\to\infty).
\end{equation*}
Note that the r.h.s. is formally obtained by term-by-term integration of the l.h.s..

Using the arguments of \cite[Ch~8.11]{cr:H49}, one concludes that there
exists an $s_0>0$, such that there is
a \emph{unique} function $F(s)$ analytic for $\Re(s)\ge s_0$
with the above asymptotic expansion. It is given by
\begin{equation}\label{Ecr:Frect}
F(s)=\Ei(s^2)\e^{s^2},
\end{equation}
where $\Ei(z)=\int_1^\infty \e^{-tz}/t\,{\rm d}t$ is the 
exponential integral. The moment generating function 
$M(t)=\mathbb E[\e^{-tX}]$ of the random variable $X$ is 
given by an inverse Laplace transform of $F(s)$,
\begin{equation*}
\int_0^\infty \e^{-st}M(t^2)t\,{\rm d}t=F(s).
\end{equation*}

Since there are effective methods for computing inverse 
Laplace transforms \cite{cr:D74}, the question arises whether
the function $F(s)$ can be easily obtained. It turns out that the 
functional equation Eq.~(\ref{Ecr:recfeq}) induces a differential 
equation for $F(s)$. This equation can be obtained in a 
mechanical way, using the method of dominant balance.

\subsubsection{Dominant balance}\index{dominant balance}

For a given functional equation, the method of dominant 
balance consists of a certain rescaling of the variables, 
such that the quantity of interest appears in the expansion 
of a rescaled variable to leading order.
The method was originally used as an heuristic tool in order 
to extract the scaling function of a polygon model 
\cite{cr:PB95} (see the following section). In the present 
framework, it is a rigorous method.

\medskip

Consider the half-perimeter and area generating
function $P(x,q)$ as a formal power series. 
The substitution $q=1-\widetilde\epsilon$ is valid,
since the coefficients of the power series $P(x,q)$ 
in $x$ are polynomials in $q$. We get the 
power series in $\widetilde\epsilon$,
\begin{equation*}
H(x,\widetilde\epsilon) =\sum_{k\ge0} (-1)^kg_k(x)\widetilde\epsilon^k.
\end{equation*}
whose coefficients $(-1)^kg_k(x)$ are power series in $x$.
The functional equation Eq.~(\ref{Ecr:recfeq}) induces an equation for
$H(x,\widetilde\epsilon)$, from which the factorial area 
moment generating functions $g_k(x)$ may be computed 
recursively. 

\smallskip

Now replace $g_k(x)$ by its expansion about $x=1$,
\begin{equation*}
g_k(x)=\sum_{l\ge0} \frac{f_{k,l}}{(1-x)^{2k+2-l}}.
\end{equation*}
Introducing $\widetilde s=1-x$, this leads to a power series 
$E(\widetilde s,\widetilde\epsilon)$ in $\widetilde\epsilon$,
\begin{equation*}
E(\widetilde s,\widetilde\epsilon) =\sum_{k\ge0} (-1)^k\left(\sum_{l\ge0} 
\frac{f_{k,l}}{\widetilde s^{2k+2-l}}\right)
\widetilde\epsilon^k,
\end{equation*}
whose coefficients are Laurent series in $\widetilde s$. As above, the
functional equation induces an equation for the power
series $E(\widetilde s,\widetilde\epsilon)$ in $\widetilde\epsilon$,
from which the expansion coefficients may be computed recursively.

\smallskip

We infer from the previous equation that
\begin{equation}\label{Ecr:rectscal}
E(s\epsilon,\epsilon^2)=\frac{1}{\epsilon^2}\sum_{l\ge0} \left(\sum_{k\ge0} 
(-1)^k\frac{f_{k,l}}{s^{2k+2-l}}\right)
\epsilon^{l}=\frac{1}{\epsilon^2}F(s,\epsilon).
\end{equation}
Write $F(s,\epsilon)=\sum_{l\ge0} F_l(s)\epsilon^l$. By construction, 
the (formal) series $F_0(s)=F(s,0)$ coincides with the asymptotic expansion of 
the desired function $F(s)$ Eq.~(\ref{Ecr:Frect}) about infinity. 

\medskip

The above example suggests a technique for computing $F_0(s)$. 
The functional equation Eq.~(\ref{Ecr:recfeq}) 
for $P(x,q)$ induces, after reparametrisation, differential equations
for the functions $F_l(s)$, from which $F_0(s)$ may be obtained
explicitly. These may be computed by first writing
\begin{equation}\label{Ecr:scalans}
P(x,q)=\frac{1}{1-q} F\left(\frac{1-x}{(1-q)^{1/2}},(1-q)^{1/2}\right),
\end{equation}
and then introducing variables $s$ and $\epsilon$, by setting 
$x=1-s\epsilon$ and $q=1-\epsilon^2$. Expand the equation to 
leading order in $\epsilon$. This yields, to order $\epsilon^0$, the 
first order differential equation 
\begin{equation*}
sF_0'(s)+2-2s^2F_0(s)=0.
\end{equation*}
The above equation translates into a recursion for the coefficients 
$f_{k,0}$, from which $f_{k,0}=k!$ can be deduced. In addition, 
the equation has a unique solution with the prescribed asymptotic 
behaviour Eqn.~(\ref{Ecr:rectscal}), which is given by 
$F_0(s)=\Ei(s^2)\e^{s^2}$.

As we will argue in the next section, Eq.~(\ref{Ecr:scalans}) 
is sometimes referred to as a \emph{scaling Ansatz}, the 
function $F(s,0)$ appears as a \emph{scaling 
function}\index{scaling function}, the functions $F_l(s)$, for $l\ge1$, 
appear as \emph{correction-to-scaling
functions}\index{correction-to-scaling function}. In our formal 
framework, where the series $F_l(s)$ are rescaled generating 
functions for the coefficients $f_{k,l}$, their derivation is rigorous.

\subsection{A general method}\label{cr:sec:general}

In the preceding two subsections, we described a
method for obtaining limit laws of counting parameters, via
a generating function approach. Since this method will be
important in the remainder of this section, we
summarise it here. Its first ingredient is based
on the so-called method of moments\index{method of moments} 
\cite[Thm~4.5.5]{cr:Ch74}.

\begin{proposition}\label{cr:theo:general}
For $m,n\in\mathbb N_0$, let real numbers $p_{m,n}$ be given.
Assume that the numbers $p_{m,n}$ asymptotically satisfy, 
for $k\in\mathbb N_0$,
\begin{equation}\label{Ecr:ampli2}
\frac{1}{k!}\sum_n (n)_k p_{m,n}\sim A_k x_c^{-m} 
m^{\gamma_k-1}\qquad (m\to\infty),
\end{equation}
where $A_k$ are positive numbers, and $\gamma_k=(k-\theta)/\phi$, 
with real constants $\theta$ and $\phi>0$.
Assume that the numbers $M_k:=A_k/A_0$ satisfy the Carleman condition
\begin{equation}\label{Ecr:growth}
\sum_{k=1}^\infty (M_{2k})^{-1/(2k)}=+\infty.
\end{equation}
Then the following conclusions hold.

\begin{itemize}
\item[{\it i)}]
For almost all $m$, the random variables $\widetilde X_m$
\begin{equation}\label{Ecr:xmt}
\mathbb P(\widetilde X_m=n)=\frac{p_{m,n}}{\sum_np_{m,n}}
\end{equation}
are well defined. We have
\begin{equation}\label{Ecr:norm}
X_m:=\frac{\widetilde X_m}{m^{1/\phi}} \stackrel{d}{\to}X,
\end{equation}
for a unique random variable $X$ with moments $M_k$, 
where $\stackrel{d}{}$ denotes convergence in distribution. 
We also have moment convergence. 
\item[{\it ii)}]
If the numbers $M_k$ satisfy for all $t\in\mathbb R$
the estimate
\begin{equation}\label{Ecr:growth2}
\lim_{k\to\infty}\frac{M_kt^k}{k!}=0,
\end{equation}
then the moment generating function $M(t)=\mathbb E[\e^{-tX}]$ of $X$ 
is an entire function. The coefficients $A_k\Gamma(\gamma_k)$ are related to 
$M(t)$ by a Laplace transform which has, for $\theta>0$, the asymptotic expansion
\begin{equation}\label{Ecr:wat}
\begin{split}
\int_0^\infty \e^{-st} &\left(\sum_{k\ge0}\frac{\mathbb E[X^k]}{k!} 
(-t^{1/\phi})^k\right)\frac{1}{t^{1-\gamma_0}} \,{\rm d}t\\
&\sim \frac{1}{A_0}\sum_{k\ge0} (-1)^k
A_k\Gamma(\gamma_k) s^{-\gamma_k} \qquad (s\to\infty).
\end{split}
\end{equation}
\end{itemize}
\end{proposition}

\begin{proof}[sketch]
A straightforward calculation using Eq.~(\ref{Ecr:ampli2}) leads to
\begin{equation*}
\frac{\mathbb E[(\widetilde X_m)_k]}{k!}
\sim \frac{A_k}{A_0} m^{k/\phi}
\qquad (m\to\infty).
\end{equation*}
This implies that the same asymptotic form holds
for the (ordinary) moments $\mathbb E[(\widetilde X_m)^k]$.
Due to the growth condition Eq.~(\ref{Ecr:growth}), the sequence 
$(M_k)$ defines a unique random variable $X$ with moments $M_k$.
Also, moment convergence of the sequence $(X_m)$ to $X$ implies 
convergence in distribution, see \cite[Thm~4.5.5]{cr:Ch74}.
Due to the growth condition Eq.~(\ref{Ecr:growth2}),
the function $M(t)$ is entire. Hence the conditions of
Watson's Lemma \cite[Sec~4.1]{cr:BH86} are satisfied, and we
obtain Eq.~(\ref{Ecr:wat}).
\qed
\end{proof}

\noindent {\bf Remarks.} 
{\it i)} The growth condition Eq.~(\ref{Ecr:growth2}) 
implies the Carleman condition Eq.~(\ref{Ecr:growth}).
All examples below have entire moment generating
functions $M(t)$.
\\
{\it ii)} If $\gamma_0<0$, a modified version of
Eq.~(\ref{Ecr:wat}) can be given, see for example 
staircase polygons below.

\medskip

Proposition \ref{cr:prop:gksing} states that
assumption Eq.~(\ref{Ecr:ampli2}) translates, at the 
level of the half-perimeter and area generating function
$P(x,q)=\sum_{m,n}p_{m,n}x^mq^n$, to a certain asymptotic
expression for the factorial moment generating functions
\begin{equation*}
g_k(x)=\frac{1}{k!}\left.\frac{\partial^k}{\partial q^k}
P(x,q)\right|_{q=1}.
\end{equation*}
Their asymptotic behaviour follows from Eq.~(\ref{Ecr:ampli2}), 
and is
\begin{equation*}
g^{(sing)}_k(x)\sim \frac{f_k}{(1-x/x_c)^{\gamma_k}}\qquad (x\nearrow x_c),
\end{equation*}
where $f_k=A_k\Gamma(\gamma_k)$. Adopting the generating function
viewpoint, the amplitudes $f_k$ determine the numbers $A_k$,
hence the moments $M_k=A_k/A_0$ of the limit distribution.
The series $F(s)=\sum_{k\ge0} (-1)^kf_k s^{-\gamma_k}$
will be of central importance in the sequel. 

\begin{definition}[Area amplitude series]\label{cr:def:fscl}
\index{area amplitude series}
Let Assumption \ref{Acr:pol} be satisfied. Assume that the 
generating function $P(x,q)=\sum_{m,n}p_{m,n}x^mq^n$ satisfies asymptotically
\begin{equation*}
\begin{split}
\left(\frac{1}{k!}\left.\frac{\partial^k}{\partial q^k} P(x,q)
\right|_{q=1}\right)^{(sing)}
\sim \frac{f_k}{(1-x/x_c)^{\gamma_k}} \qquad (x\nearrow x_c),\\
\end{split}
\end{equation*}
with exponents $\gamma_k\notin \{0,-1,-2,\ldots\}$. Then, the formal series
\begin{displaymath}
F(s) = \sum_{k\ge0}(-1)^k\frac{f_k}{s^{\gamma_k}}
\end{displaymath}
is called the \emph{area amplitude series}.
\end{definition}

\noindent {\bf  Remarks.} 
{\it i)} Proposition~\ref{cr:theo:general} states that the area 
amplitude series appears in the asymptotic expansion about infinity of
a Laplace transform of the moment generating function of the area limit 
distribution. The probability distribution of the
limiting area distribution is related to $F(s)$ by a double
Laplace transform.\\
{\it ii)} For typical polygon models, all derivatives of $P(x,q)$ w.r.t.~$q$,
evaluated at $q=1$, exist and have the same radius of convergence, 
see Proposition~\ref{cr:prop:pol}. Typical polygon models do have
factorial moment generating functions of the above form, see the
examples below.

\medskip

The second ingredient of the method consists in applying the method
of dominant balance. As described above, this may result in a 
differential equation (or in a difference equation \cite{cr:RG01}) 
for the function $F(s)$. Its applicability
has to be tested for each given type of functional equation.
Typically, it can be applied if the factorial area moment 
generating functions $g_k(x)$ Eq.~(\ref{Ecr:gengk}) have,
for values $x<x_c$, a local expansion about $x=x_c$ of the 
form
\begin{equation*}
g^{(sing)}_k(x)=\sum_{l\ge0} \frac{f_{k,l}}{(1-x/x_c)^{\gamma_{k,l}}},
\end{equation*}
where $\gamma_{k,l}=(k-\theta_l)/\phi$ and $\theta_{l+1}>\theta_l$. 
If a transfer theorem such as Lemma~\ref{Ecr:transfergx} applies,
then the differential equation for $F(s)$ induces a recurrence for the 
moments of the limit distribution. If the differential equation 
can be solved in closed form, inverse Laplace transform techniques 
may be applied in order to obtain explicit expressions for the moment
generating function and the probability density. Also, higher order 
corrections to the limiting behaviour may be analysed, by studying
the functions $F_l(s)$, for $l\ge1$. See \cite{cr:Rich02} for examples.

\subsection{Further examples}

Using the general method as described above, area limit laws for the
other exactly solved polygon models can be derived. A model with the same
area limit law as rectangles is convex polygons\index{convex polygons}, 
compare \cite{cr:Rich02}. We will discuss some classes of polygon 
models with different area limit laws.

\subsubsection{Ferrers diagrams}\index{Ferrers diagrams}

In contrast to the previous example, the limit distribution
of area of Ferrers diagrams is concentrated.

\begin{proposition}\label{cr:prop:ferr}
The area random variable $\widetilde X_m$ of Ferrers diagrams
has mean $\mu_m\sim m^2/8$. The normalised random variables 
$X_m$ Eq.~(\ref{Ecr:norm}) converge in distribution
to a random variable with density $p(x)=\delta(x-1/8)$.
\end{proposition}

\noindent {\bf  Remark.} It should be noted that the above
convergence statement already follows from the concentration property 
$\lim_{m\to\infty}\sigma_m/\mu_m=0$, with $\sigma_m^2\sim m^3/48$ the variance
of $X_m$,  by an explicit analysis of the first three factorial
moment generating functions.
(By Chebyshev's inequality, the concentration property implies convergence 
in probability, which in turn implies convergence in distribution.) For 
illustrative purposes, we follow a different route via the moment 
method in the following proof.

\begin{proof}
Ferrers diagrams, counted by half-perimeter and area, satisfy the linear
$q$-difference equation \cite[Eq~(5.4)]{cr:Rich02}
\begin{equation*}
P(x,q)=\frac{qx^2}{(1-qx)^2}P(qx,q)+\frac{qx^2}{(1-qx)^2}.
\end{equation*}
The perimeter generating function $g_0(x)=x^2/(1-2x)$ is obtained
by setting $q=1$ in the above equation. Hence $x_c=1/2$. Using 
the functional equation, it can be shown by induction on $k$ that all 
area moment generating functions $g_k(x)$ are rational in $g_0(x)$ 
and its derivatives. Hence all $g_k(x)$ are rational functions. Since the 
area of a polygon grows at most quadratically with the perimeter, we 
have a bound on the exponent, $\gamma_{k}\le2k+1$, of the leading 
singular part of $g_k(x)$. Given this bound, the method of dominant 
balance can be applied. We set
\begin{equation*}
P(x,q)=\frac{1}{(1-q)^{\frac{1}{2}}}F\left(\frac{1-2x}
{(1-q)^{\frac{1}{2}}},(1-q)^{\frac{1}{2}} \right),
\end{equation*}
and introduce new variables $s$ and $\epsilon$ by $q=1-\epsilon^2$ and 
$2x=1-s\epsilon$. Then an expansion of the functional equation yields,
to order $\epsilon^0$, the ODE of first order $F'(s)=4sF(s)-1$, whose 
unique solution with the prescribed asymptotic behaviour is 
\begin{equation*}
F(s)=\sqrt{\frac{\pi}{8}}\erfc\left(\sqrt{2}s\right)\e^{2s^2}.
\end{equation*}
It can be inferred from the differential equation that all coefficients 
in the asymptotic expansion of $F(s)$ at infinity are nonzero. Hence,
 the above exponent bound is tight. It can be inferred from the functional
equation by induction on $k$ that each $g_k(x)$ is a Laurent polynomial 
about $x_c=1/2$. Thus, Lemma~\ref{Ecr:transfergx} applies, and we obtain the 
moment generating function of the corresponding random variable
Eq.~(\ref{Ecr:norm}) as $M(s)=\exp(-s/8)$. This is readily recognised
as the moment generating function of a probability distribution concentrated
at $x=1/8$.
\qed
\end{proof}

A sequence of random variables, which satisfies the concentration 
property, often leads to a Gaussian limit law, after centering and 
suitable normalisation. This is also the case for Ferrers diagrams.

\begin{theorem}[\cite{cr:S08b}]
The area random variable $\widetilde X_m$ of Ferrers diagrams
has mean $\mu_m\sim m^2/8$ and variance 
$\sigma_m^2\sim m^3/48$. The centred and normalised random
variables
\begin{equation}\label{Ecr:normcent}
X_m=\frac{\widetilde X_m-\mu_m}{\sigma_m},
\end{equation}
converge in distribution to a Gaussian random variable. \qed
\end{theorem}
\noindent {\bf  Remarks.} 
{\it i)} It is possible to prove this result by the
method of dominant balance. The idea of proof consists
in studying the functional equation of the generating function
for the ``centred coefficients'' $p_{m,n}-\mu_m$.\\
{\it ii)}
The above arguments can also be 
applied to stack polygons\index{stack polygons} to yield the concentration
property and a central limit theorem.

\subsubsection{Staircase polygons}\label{cr:sec:stair}\index{staircase polygons}

The limit law of area of staircase polygons is 
the Airy distribution\index{Airy distribution}. This distribution 
(see \cite{cr:FL01} and the survey \cite{cr:J07}) is conveniently 
defined via its moments.

\begin{definition}[Airy distribution \cite{cr:FL01}]\label{cr:def:Airy}
The random variable $Y$ is said to be Airy distributed if
\begin{equation*}
\frac{\mathbb E[Y^k]}{k!} = \frac{\Gamma(\gamma_0)}{\Gamma(\gamma_k)} 
\frac{\phi_k}{\phi_0},
\end{equation*}
where $\gamma_k=3k/2-1/2$, and the numbers $\phi_k$ satisfy, for $k\ge1$, 
the quadratic recurrence
\begin{equation*}
\gamma_{k-1}\phi_{k-1}+\frac{1}{2}\sum_{l=0}^k \phi_l \phi_{k-l}=0,
\end{equation*}
with initial condition $\phi_0=-1$.
\end{definition}

\noindent {\bf Remarks (\cite{cr:FL01,cr:KMM07}).} {\it i)} The
first moment is $\mathbb E[Y]=\sqrt{\pi}$. The 
sequence of moments can be shown to satisfy the Carleman condition. 
Hence the distribution is uniquely determined by its moments.\\
{\it ii)} The numbers $\phi_k$ appear in the asymptotic
expansion of the logarithmic derivative of the
Airy function at infinity,
\begin{equation*}
\frac{{\rm d}}{{\rm d}s}\log\Ai(s)\sim \sum_{k\ge0} 
(-1)^k \frac{\phi_k}{2^k} s^{-\gamma_k} \qquad (s\to\infty),
\end{equation*}
where $\Ai(x)=\frac{1}{\pi}\int_0^\infty \cos(t^3/3+tx)\,{\rm d}t$
is the Airy function.\\
{\it iii)} Explicit expressions for the numbers $\phi_k$ are known \cite{cr:KMM07}.
They are, for $k\ge1$, given by
\begin{displaymath}
\phi_k=2^{k+1}\frac{3}{4\pi^2}\int_0^\infty \frac{x^{3(k-1)/2}}
{\Ai(x)^2+\Bi(x)^2}\,{\rm d}x,
\end{displaymath}
where $\Bi(z)$ is the second standard solution of the
Airy differential equation $f''(z)-zf(z)=0$.\\
{\it iv)} The Airy distribution appears in a variety
of contexts \cite{cr:FL01}. In particular, the random 
variable $Y/\sqrt{8}$ describes the law of the area of 
a Brownian excursion. See also \cite{cr:MC05} for an overview 
from a physical perspective.

\medskip

Explicit expressions have been derived for the moment
generating function of the Airy distribution and for 
its density.

\begin{fact}[\cite{cr:D83,cr:L84,cr:T91,cr:FL01}]\label{cr:prop:Airy}
The moment generating function $M(t)=\mathbb E[\e^{-tY}]$ of 
the Airy distribution satisfies the modified Laplace transform 
\begin{equation}\label{Ecr:main}
\frac{1}{\sqrt{2\pi}}\int_0^\infty (\e^{-st}-1)M(2^{-3/2}t^{3/2}) \frac{1}{t^{3/2}}{\rm d}t 
= 2^{1/3}\left( \frac{\Ai'(2^{1/3}s)}{\Ai(2^{1/3}s)}-
\frac{\Ai'(0)}{\Ai(0)}\right).
\end{equation}
The moment generating function $M(t)$ is given explicitly by
\begin{equation*}
M(2^{-3/2}t)=\sqrt{2\pi}t\sum_{k=1}^\infty \exp\left( -\beta_k t^{2/3}2^{-1/3}\right),\\
\end{equation*}
where the numbers $-\beta_k$ are the zeros of the Airy function. 
Its density $p(x)$ is given explicitly by
\begin{equation*}
2^{3/2}p(2^{3/2}x)=\frac{2\sqrt{6}}{x^2}\sum_{k=1}^\infty \e^{-v_k}v_k^{2/3}U
\left(-\frac{5}{6},\frac{4}{3}; v_k\right),
\end{equation*}
where $v_k=2\beta_k^3/(27x^2)$ and $U(a,b,z)$ is the
confluent hypergeometric function.\qed
\end{fact}

\noindent {\bf Remarks.} {\it i)} The confluent hypergeometric
function $U(a,b;z)$ is defined as \cite{cr:AS73}
\begin{equation*}
U(a,b;z)=\frac{\pi}{\sin\pi b}\left( \frac{{}_1F_1[a,b;z]}{\Gamma(1+a-b)\Gamma(b)}-
\frac{z^{1-b}{}_1F_1[1+a-b,2-b;z]}{\Gamma(a)\Gamma(2-b)}\right),
\end{equation*}
where ${}_1F_1[a;b;z]$ is the hypergeometric function
\begin{equation*}
{}_1F_1[a;b;z]=1+\frac{a}{b}\frac{z}{1!}+
\frac{a(a+1)}{b(b+1)}\frac{z^2}{2!}+\ldots
\end{equation*}
{\it ii)} The moment generating function and
its density are obtained by two consecutive inverse Laplace transforms
of Eq.~(\ref{Ecr:main}), see \cite{cr:L85,cr:L86} and \cite{cr:T91,cr:JK83}.\\
{\it iii)} In the proof of the following theorem, we will derive 
Eq.~(\ref{Ecr:main}) using the model of staircase polygons. This
shows, in particular, that the coefficients $\phi_k$ appear in the
asymptotic expansion of the Airy function.

\begin{theorem}
The normalised area random variables $X_m$ of staircase
polygons Eq.~(\ref{Ecr:norm}) satisfy
\begin{equation*}
\frac{X_m}{\sqrt{\pi}/4} \stackrel{d}{\longrightarrow} \frac{Y}{\sqrt{\pi}} \qquad (m\to\infty),
\end{equation*}
where $Y$ is Airy distributed according to 
Definition \ref{cr:def:Airy}. We also have moment convergence.
\end{theorem}

\noindent {\bf Remark.} Given the functional equation of the 
half-perimeter and area generating function of staircase polygons,
\begin{equation}\label{Ecr:stairfeq}
P(x,q)=\frac{x^2q}{1-2xq-P(qx,q)}
\end{equation}
(see \cite{cr:Rich06} for a recent derivation), this result is a special
case of Theorem \ref{theo:qfuncAiry} below, which
is stated in \cite{cr:Duchon99}. 

\begin{proof}
We use the method of dominant balance. From the functional equation  
Eq.~(\ref{Ecr:stairfeq}), we infer $g_0(x)=1/4+\sqrt{1-4x}/2+(1-4x)/4$. 
Hence $x_c=1/4$. The structure of the functional equation implies that 
all functions $g_k(x)$ can be written as Laurent series in $s=\sqrt{1-4x}$, 
see also Proposition \ref{cr:prop:gk} below. Explicitly, we get $g_1(x)=x^2/(1-4x)$. 
This suggests $\gamma_k=(3k-1)/2$. An upper bound of this form on the exponent 
$\gamma_k$ can be derived without too much effort from the functional equation, 
by an application of Faa di Bruno's formula, see also \cite[Prop~(4.4)]{cr:Rich05}. 
Thus, the method of dominant balance can be applied. We set
\begin{equation*}
P(x,q)=\frac{1}{4}+(1-q)^{1/3} F\left(\frac{1-4x}{(1-q)^{2/3}},(1-q)^{1/3}\right)
\end{equation*}
and introduce variables $s,\epsilon$ by $4x=1-s\epsilon^2$ and 
$q=1-\epsilon^3$. In the above equation,
we excluded the constant $1/4=:P^{(reg)}(x,q)$, since it does not
contribute to the moment asymptotics. Expanding the functional 
equation to order $\epsilon^2$ gives the Riccati equation 
\begin{equation}\label{Ecr:Ric}
F'(s)+4F(s)^2-s=0.
\end{equation}
It follows that the coefficients $f_k$ of $F(s)$ satisfy, for $k\ge1$, the 
quadratic recursion
\begin{equation*}
\gamma_{k-1}f_{k-1}+4\sum_{l=0}^k f_l f_{k-l}=0,
\end{equation*}
with initial condition $f_0=-1/2$. A comparison with the definition
of the Airy distribution shows that $\phi_k=2^{2k+1}f_k$. Using the
closure properties of $\Delta$-regular functions, it can be 
inferred from the functional equation that (the analytic continuation of) 
each factorial moment generating function $g_k(x)$ is $\Delta$-regular, 
with $x_c=1/4$, see also Proposition \ref{cr:prop:gk} below.
Hence the transfer theorem Lemma \ref{Ecr:transfergx} can be applied.
We obtain $4X_m\stackrel{d}{\to} Y$ in distribution and for moments, 
where $Y$ is Airy distributed.
\qed
\end{proof}

\noindent {\bf Remarks.} 
{\it i)} The unique solution $F(s)$ of the differential 
equation in the above proof Eq.~(\ref{Ecr:Ric}), satisfying the
prescribed asymptotic behaviour, is given by
\begin{equation}\label{Ecr:stairscal}
F(s)=\frac{1}{4} \frac{{\rm d}}{{\rm d}s} \log\Ai(4^{1/3}s).
\end{equation}
The moment generating function $M(t)$ of the limiting random variable
$X=\lim_{m\to\infty}X_m$ is related to the function $F(s)$ via
the modified Laplace transform
\begin{equation*}
\int_0^\infty (\e^{-st}-1)M(t^{3/2}) \frac{1}{t^{3/2}}\,{\rm d}t = 4\sqrt{\pi}(F(s)-F(0)),
\end{equation*}
where the modification has been introduced in order to ensure a finite
integral about the origin. This result relates the above proof to 
Proposition \ref{cr:prop:Airy}.\\
{\it ii)} The method of dominant balance can be used to obtain
corrections $F_l(s)$ to the limiting behaviour \cite{cr:Rich02}.

\medskip

\begin{figure}[htb]
\begin{center}
\begin{minipage}[b]{0.95\textwidth}
\center{\includegraphics[width=10cm]{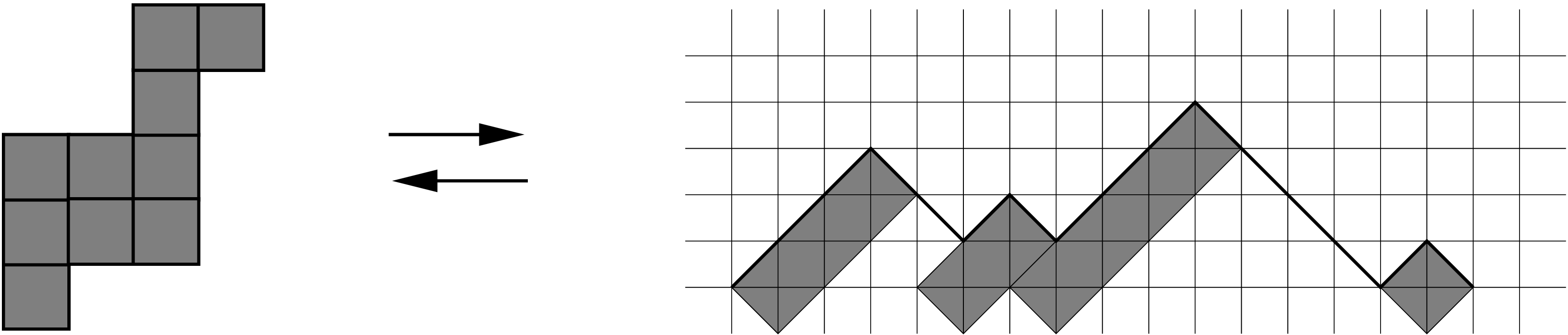}}
\end{minipage}
\end{center}
\caption{\cite{cr:Rich06}\label{Fcr:pol}
\small A combinatorial bijection between staircase polygons and Dyck paths 
\cite{cr:DV84,cr:S99}. Column heights of a polygon correspond to peak heights of a path.}
\end{figure}

The fact that the area law of staircase polygons is, up to
normalisation, the same as that of the area under a Brownian
excursion, suggests that there might be a combinatorial explanation. 
Indeed, as is well known, there is a bijection \cite{cr:DV84,cr:S99} between 
staircase polygons and Dyck paths, a discrete version of Brownian
excursions \cite{cr:A92}, see figure \ref{Fcr:pol} \cite{cr:Rich06}. Within 
this bijection, the polygon area corresponds to the sum of peak heights
of the Dyck path, but not to the area below the Dyck path.
For more about this connection, see the remark at
the end of the following subsection.

\subsection{$q$-difference equations}

All polygon models discussed above have
an algebraic perimeter generating function.
Moreover, their half-perimeter and area
generating function satisfies a functional 
equation of the form
\begin{equation*}
P(x,q)=G(x,q,P(x,q),P(qx,q)),
\end{equation*}
for a real polynomial $G(x,q,y_0,y_1)$. Since, under mild
assumptions on $G$, the equation reduces to an algebraic 
equation for $P(x,1)$ in the limit $q\to1$, it may be viewed 
as a ``deformation'' of an algebraic equation. In this 
subsection, we will analyse equations of this type
at the special point $(x,q)=(x_c,1)$, where $x_c$
is the radius of convergence of $P(x,1)$.
It will appear that the methods used in the
above examples also can be applied to this more general case.

The above equation falls into the class of
\emph{$q$-difference equations} \cite{cr:DRSZ03}. While 
particular examples appear in combinatorics in a number
of places, see e.g.~\cite{cr:FS06}, the asymptotic behaviour 
of equations of the above form seems to have been 
systematically studied initially in \cite{cr:Duchon99,cr:Rich02}.
The study can be done in some generality, e.g., also for
non-polynomial power series $G$, for 
replacements more general than $x\mapsto qx$, and for
multivariate generalisations, see \cite{cr:Rich05} and 
\cite{cr:Duchon99}. For simplicity, we will concentrate on polynomial
$G$, and then briefly discuss generalisations. Our 
exposition closely follows \cite{cr:Rich05, cr:Rich02}.

\subsubsection{Algebraic $q$-difference equations}\index{$q$-difference equation}

\begin{definition}[Algebraic $q$-difference equation \cite{cr:Duchon99,cr:Rich02}]\label{cr:def:feq}
An algebraic $q$-difference equation is an equation of the form
\begin{equation}\label{Ecr:qalgfe}
P(x,q)=G(x,q,P(x,q),P(qx,q),\ldots,P(q^Nx,q)),
\end{equation}
where $G(x,q,y_0,y_1,\ldots,y_N)$ is 
a complex polynomial. We require that 
\begin{equation*}
G(0,q,0,0,\ldots,0)\equiv0,\qquad 
\frac{\partial G}{\partial y_k}(0,q,0,0,\ldots,0)\equiv0
\qquad (k=0,1,\ldots,N).
\end{equation*}
\end{definition}

\noindent {\bf  Remarks.} {\it i)} See \cite{cr:DRSZ03} 
for an overview of the theory of $q$-difference equations. 
As $q$ approaches unity, the above equation reduces to an 
algebraic equation.\\
{\it ii)} Asymptotics for solutions of algebraic $q$-difference 
equations have been considered in \cite{cr:Duchon99}. The above definition 
is a special case of \cite[Def~2.4]{cr:Rich05}, where a
multivariate extension is considered, and where
$G$ may be non-polynomial. Also, replacements more
general than $x\mapsto f(q)x$ are allowed. Such equations
are called $q$-functional equations in \cite{cr:Rich05}. The 
results presented below apply {\it mutatis mutandis} 
also to $q$-functional equations.

\medskip

The algebraic $q$-difference equation in Definition~\ref{cr:def:feq}
uniquely defines a (formal) power series $P(x,q)$ satisfying
$P(0,q)\equiv0$. This is shown by analysing the implied recurrence 
for the coefficients $p_m(q)$ of $P(x,q)=\sum_{m>0} p_m(q)x^m$, 
see also \cite[Prop~2.5]{cr:Rich05}. In fact, $p_m(q)$ is a polynomial 
in $q$. The growth of its degree in $m$ is not larger than
$cm^2$ for some positive constant $c$, hence the counting parameters 
are rank 2 parameters \cite{cr:Duchon99}. In our situation, 
such a bound holds, since the area of a polygon grows at 
most quadratically with its perimeter. 

\smallskip

From the preceding discussion, it follows that the factorial
moment generating functions
\begin{equation*}
g_k(x)=\frac{1}{k!}\left.\frac{\partial^k}{\partial q^k}P(x,q)\right|_{q=1}
\end{equation*}
are well-defined as formal power series. In fact,
they can be recursively determined from the $q$-difference 
equation by implicit differentiation, as a consequence of the
following proposition.

\begin{proposition}[{\cite{cr:Rich02,cr:Rich05}}]
Consider the derivative of order $k>0$ of an algebraic $q$-difference equation
Eq.~\eqref{Ecr:qalgfe} w.r.t. $q$, evaluated at $q=1$. It is linear in $g_k(x)$, and its 
r.h.s. is a complex polynomial in the power series $g_l(x)$ and its derivatives up to 
order $k-l$, where $l=0,\ldots,k$.
\qed
\end{proposition}

\noindent {\bf  Remarks.} {\it i)} This statement can be shown
by analysing the $k$-th derivative of the $q$-difference equation,
using Faa di Bruno's formula \cite{cr:CS96}.\\
{\it ii)} It follows that every function $g_k(x)$ is rational in $g_l(x)$ and its 
derivatives up to order $k-l$, where $0\le l<k$. Since $G$ is a 
polynomial, $g_k(x)$ is \emph{algebraic}, by the closure 
properties of algebraic functions.

\medskip

We discuss analytic properties of the (analytic continuations of the) 
factorial moment generating functions $g_k(x)$. These are determined
by the analytic properties of $g_0(x)=P(x,1)$. 
We discuss the case of a square-root singularity of 
$P(x,1)$, which often occurs for combinatorial structures, and
which is well studied, see e.g.~\cite[Thm~10.6]{cr:O95} or \cite[Ch~VII.4]{cr:FS06}. 
Other cases may be treated similarly. We make the following assumption:

\begin{ass}\label{cr:ass:qdiff}
The $q$-difference equation in Definition \ref{cr:def:feq} 
has the following properties:
\begin{itemize}
\item[{\it i)}] All coefficients of the polynomial 
$G(x,q,y_0,y_1,\ldots,y_N)$ are non-negative.
\item[{\it ii)}] The polynomial $Q(x,y):=G(x,1,y,y,\ldots,y)$ satisfies 
$Q(x,0)\not\equiv0$ and has degree at least two in $y$.
\item[{\it iii)}] $P(x,1)=\sum_{m\ge1} p_m x^m$ is \emph{aperiodic},
i.e., there exist indices $1\le i<j<k$ such that $p_ip_jp_k\ne0$, 
while $\gcd(j-i,k-i)=1$.
\end{itemize}
\end{ass}

\noindent {\bf  Remarks.} {\it i)} The positivity assumption
is natural for combinatorial constructions. There are, however,
$q$-difference equations with negative coefficients, which arise 
from systems of $q$-difference equations with non-negative 
coefficients by reduction. Examples are convex polygons 
\cite[Sec~5.4]{cr:Rich02} and directed convex polygons, see below.\\ 
{\it ii)} Assumptions $i)$ and $ii)$ result in a square-root 
singularity as the dominant singularity of $P(x,1)$.\\
{\it iii)} Assumption $iii)$ implies that there is only one singularity 
of $P(x,1)$ on its circle of convergence. Since
$P(x,1)$ has non-negative coefficients only, it occurs on the positive
real half-line. The periodic case can be treated by a straightforward
extension \cite{cr:FS06}.

\medskip

An application of the (complex) implicit function theorem ensures that $P(x,1)$ is
analytic at the origin. It can be analytically continued, 
as long as the defining algebraic equation remains invertible.
Together with the positivity assumption, one can conclude 
that there is a number $0<x_c<\infty$, 
such that the analytic continuation of $P(x,1)$ satisfies 
$y_c=\lim_{x\nearrow x_c}P(x,1)<\infty$, with
\begin{equation*}
Q(x_c,y_c)=y_c, \quad \left.\frac{\partial}{\partial y}Q(x_c,y)\right|_{y=y_c}=1.
\end{equation*}
With the positivity assumption on the coefficients,
it follows that
\begin{equation}\label{Ecr:BC}
B:=\frac{1}{2}\left.\frac{\partial^2}{\partial y^2}Q(x_c,y)\right|_{y=y_c}>0, \quad
C:=\left.\frac{\partial}{\partial x}Q(x,y_c)\right|_{x=x_c}>0.
\end{equation}

These conditions characterise the singularity of $P(x,1)$ 
at $x=x_c$ as a square-root. It can be shown that there exists a locally convergent 
expansion of $P(x,1)$ about $x=x_c$, and that $P(x,1)$ is analytic for $|x|<x_c$. 
We have the following result. Recall that a function $f(z)$ is 
$\Delta$-regular if it is analytic in the 
indented disc $\Delta=\{z:|z|\le x_c+\eta, 
|\Arg(z-x_c)|\ge\phi\}$ for some $\eta>0$ and some 
$\phi$, where $0<\phi<\pi/2$. 

\begin{proposition}[\cite{cr:O95,cr:FS06,cr:Rich05}]\label{cr:prop:squareroot}
Given Assumption \ref{cr:ass:qdiff}, the power series $P(x,1)$ is
analytic at $x=0$, with radius of convergence $x_c$. Its analytic
continuation is $\Delta$-regular, with a square-root singularity 
at $x=x_c$ and a local Puiseux expansion
\begin{equation*}
P(x,1)=y_c+\sum_{l=0}^\infty f_{0,l}(1-x/x_c)^{1/2+l/2},
\end{equation*}
where $y_c=\lim_{x\nearrow x_c}P(x,1)<\infty$ and $f_{0,0}=-\sqrt{x_c\,C/B}$, 
for constants $B>0$ and $C>0$ as in Eq.~(\ref{Ecr:BC}). The numbers 
$f_{0,l}$ can be recursively determined from the $q$-difference equation.
\qed
\end{proposition}

The asymptotic behaviour of $P(x,1)=g_0(x)$ carries over to the 
factorial moment generating functions $g_k(x)$.

\begin{proposition}[{\cite{cr:Rich05}}]\label{cr:prop:gk}
Given Assumption \ref{cr:ass:qdiff}, all factorial moment generating
functions $g_k(x)$ are, for $k\ge1$, analytic at $x=0$, with radius of 
convergence $x_c$. Their analytic continuations are $\Delta$-regular, with
local Puiseux expansions
\begin{equation*}
g_k(x)=\sum_{l=0}^\infty f_{k,l}(1-x/x_c)^{-\gamma_k+l/2},
\end{equation*}
where $\gamma_k=3k/2-1/2$.
The numbers $f_{k,0}=f_k$ are, for $k\ge2$, characterised by the recursion
\begin{equation*}
\gamma_{k-1}f_{k-1}+\frac{1}{4f_1}\sum_{l=0}^k f_lf_{k-l}=0,
\end{equation*}
and the numbers $f_0<0$ and $f_1>0$ are given by
\begin{equation}\label{Ecr:consts}
f_0=-\sqrt{\frac{Cx_c}{B}}, \qquad 4f_1=\frac{\sum_{k=1}^N k\frac{\partial G}{\partial y_k}
(x_c,1,y_c,y_c,\ldots,y_c)}{B},
\end{equation}
for constants $B>0$ and $C>0$ as in Eq.~(\ref{Ecr:BC}).
\qed
\end{proposition}

\noindent {\bf  Remarks.} {\it i)} This result can be obtained
by a direct analysis of the $q$-difference equation, applying
Faa di Bruno's formula, see also \cite[Sec~2.2]{cr:Rich02}.\\
{\it ii)} Alternatively, it can be obtained by applying the method
of dominant balance to the $q$-difference equation. To this
end, one notes that all functions $g_k(x)$ are Laurent
series in $\sqrt{1-x/x_c}$, and that their leading exponents
are bounded from above by $\gamma_k$. (An upper bound 
on an exponent is usually easier to obtain than its exact value, 
since cancellations can be ignored). With these two ingredients, 
the method of dominant balance, as described above, can be 
applied. The differential equation of the function $F(s)$ then
translates, via a transfer theorem, into the above recursion 
for the coefficients. See \cite[Sec~5]{cr:Rich05}.

\medskip

The above result can be used to infer the limit distribution
of area, along the lines of Section~\ref{cr:sec:general}.

\begin{theorem}[{\cite{cr:Duchon99,cr:Rich05}}]\label{theo:qfuncAiry}
Let Assumption \ref{cr:ass:qdiff} be satisfied.
For the solution of an algebraic $q$-difference equation 
$P(x,q)=\sum_{m,n}p_{m,n}x^mq^n$,
let $\widetilde X_m$ denote the random variable
\begin{equation*}
\mathbb P(\widetilde X_m=n)=\frac{p_{m,n}}{\sum_n p_{m,n}}
\end{equation*}
(which is well-defined for almost all $m$). The mean of $\widetilde X_m$
is given by
\begin{displaymath}
\mathbb E[\widetilde X_m]\sim 2\sqrt{\pi}\frac{f_1}{|f_0|}m^{3/2}\qquad (m\to\infty),
\end{displaymath}
where the numbers $f_0$ and $f_1$ are given in Eq.~(\ref{Ecr:consts}).
The sequence of normalised random variables $X_m$ converges in distribution,
\begin{displaymath}
X_m=\frac{\widetilde X_m}{\mathbb E[\widetilde X_m]}
\stackrel{d}{\longrightarrow} \frac{Y}{\sqrt{\pi}} \qquad (m\to\infty),
\end{displaymath}
where $Y$ is Airy distributed according to 
Definition~\ref{cr:def:Airy}. We also have moment 
convergence.
\qed
\end{theorem}

\noindent {\bf  Remarks.} {\it i)} An explicit calculation 
shows that $\phi_k=|f_0|^{-1}\left(\frac{|f_0|}{2f_1}\right)^kf_k$. 
Together with Proposition \ref{cr:prop:gk}, the claim of the proof 
follows by standard reasoning, as in the examples above.\\
{\it ii)} The above theorem appears 
in \cite[Thm~3.1]{cr:Duchon99}, together with an indication of the
arguments of a proof. [There is a misprint in the definition 
of $\gamma$ in \cite[Thm~3.1]{cr:Duchon99}. In our notation $\gamma=4Bf_1$.]
Within the more general setup of $q$-functional equations, the 
theorem is a special case of \cite[Thm~1.5]{cr:Rich05}.\\
{\it iii)} The above theorem is a kind of central limit theorem
for combinatorial constructions, since the Airy distribution
arises under natural assumptions for a large class of
combinatorial constructions. For a connection to certain Brownian
motion functionals, see below.

\subsubsection{$q$-functional equations and other extensions}

We discuss extensions of the above result. Generically,
the dominant singularity of $P(x,1)$ is a square-root.  The
case of a simple pole as dominant singularity, which
generalises the example of Ferrers diagrams, has 
been discussed in \cite{cr:Rich02}. Under weak assumptions, 
the resulting limit distribution of area is concentrated.
Other singularities can also be analysed, as shown in the
examples of rectangles above and of directed convex 
polygons in the following subsection. Compare also 
\cite{cr:RG01}.

The case of non-polynomial $G$ can be discussed along the
same lines, with certain assumptions on the analyticity 
properties of the series $G$. In the undeformed
case $q=1$, it is a classical result \cite[Ch VII.3]{cr:FS06}
that the generating function has a square-root as
dominant singularity, as in the polynomial case.
One can then argue along the above lines that an
Airy distribution emerges as the limit law of the deformation
variable \cite[Thm~1.5]{cr:Rich05}. Such an extension
is relevant, since prominent combinatorial models,
such as the Cayley tree generating function, fall
into that class. See also the discussion of self-avoiding
polygons below.

The above statements also remain valid for more general
classes of replacements $x\mapsto qx$, e.g.,
for replacements $x\mapsto f(q)x$, where $f(q)$ is analytic
for $0\le q\le1$, with non-negative series coefficients about
$q=0$.
More interestingly, the idea of introducing a $q$-deformation may be
iterated \cite{cr:Duchon99}, leading to equations such as
\begin{equation}
P(x,q_1,\ldots,q_M)=G(x,P(xq_1\cdot\ldots\cdot q_M,q_1q_2\cdot\ldots\cdot q_M,
q_2q_3\cdot\ldots\cdot q_M,\ldots,q_M)).
\end{equation}
The counting parameters corresponding to $q_k$ are rank $k+1$ 
parameters, and limit distributions for
such quantities have been derived for some types of
singularities \cite{cr:N03a,cr:N03b,cr:Rich06}. There is a
central limit result for the generic case of a 
square-root singularity \cite{cr:Rich05}. This generalisation 
applies to counting parameters, which decompose
linearly under a combinatorial construction. These results
can also be obtained by an alternative method, which 
generalises to non-linear parameters, see \cite{cr:Janson03}.

The case where the limit $q$ to unity in a $q$-difference
equation is not algebraic, has not been discussed. For
example, if $G(x,q,P(x,q),P(qx,q))=0$ for some polynomial
$G$, the limit $q$ to unity might lead to an algebraic
differential equation for $P(x,1)$. This may be
seen by noting that
\begin{displaymath}
\lim_{q\to1}\frac{f(x)-f(qx)}{(1-q)}=xf'(x),
\end{displaymath}
for $f(x)$ differentiable at $x$.
Such equations are possibly related to polygon models
such as three-choice polygons \cite{cr:GJ06a} or punctured staircase
polygons \cite{cr:GJ06b}. Their perimeter generating function
is not algebraic, hence the models do not satisfy an algebraic 
$q$-difference equation as in Definition \ref{cr:def:feq}.

\subsubsection{A stochastic connection}

Lastly, we indicate a link to Brownian
motion, which appears in \cite{cr:T91,cr:T95} and 
was further developed in \cite{cr:N03a,cr:N03b,cr:Rich05,cr:Rich06}. 
As we saw in Section~\ref{cr:sec:general}, limit distributions
can, under certain conditions, be characterised by a certain 
Laplace transform of their moment generating functions. This 
approach, which arises naturally from the viewpoint of 
generating functions, can be applied to discrete versions 
of Brownian motion, excursions, bridges or meanders. Asymptotic 
results are results for the corresponding stochastic objects.
In fact, distributions of some functionals of
Brownian motion have apparently first been obtained
using this approach \cite{cr:T91,cr:T95}.

Interestingly, a similar characterisation appears
in stochastics for functionals of Brownian motion, 
via the Feynman-Kac formula. For example, Louchard's 
formula \cite{cr:L84} relates the logarithmic derivate 
of the Airy function to a certain Laplace transform 
of the moment generating function of the law of the 
Brownian excursion area. Distributions of functionals 
of Brownian motion can also be obtained by a path 
integral approach, see \cite{cr:Majumdar05} for a recent overview.

The discrete approach provides an alternative method
for obtaining information about distributions of
certain functionals of Brownian motion. For such functionals,
it provides an alternative proof of Louchard's formula 
\cite{cr:N03a,cr:N03b}. It leads, via the method of dominant balance, 
quite directly to moment recurrences for the underlying distribution.
These have been studied in the case of rank $k$ parameters 
for discrete models of Brownian motion. In particular, they 
characterise the distributions of integrals over $(k-1)$-th 
powers of the corresponding stochastic objects 
\cite{cr:N03a,cr:N03b,cr:Rich05,cr:Rich06}. Such results have apparently not 
been previously derived using stochastic methods. The generating
function approach can also be applied to classes of
$q$-functional equations with singularities different 
from those connected to Brownian motion. For a related
generalisation, see \cite{cr:BJ06}.

\emph{Vice versa}, results and techniques from
stochastics can be (and have been) analysed in 
order to study asymptotic properties of polygons.
An example is the contour process of simply generated 
trees \cite{cr:Gitt99}, which asymptotically describes the 
area of a staircase polygon. See also 
\cite{cr:L96,cr:L97,cr:L99,cr:LabaMarc07}.

\subsection{Directed convex polygons}\index{directed convex polygons}

We show that the limit law of area of
directed convex polygons in the uniform fixed 
perimeter ensemble is that of the area of the 
Brownian meander.

\begin{fact}[{\cite[Thm~2]{cr:T95}}]\label{cr:def:mean}
The random variable $Z$ of area of the Brownian 
meander\index{Brownian meander} is characterised by 
\begin{equation*}
\frac{\mathbb E[Z^k]}{k!} = \frac{\Gamma(\alpha_0)}{\Gamma(\alpha_k)} 
\frac{\omega_k}{\omega_0}\frac{1}{2^{k/2}},
\end{equation*}
where $\alpha_k=3k/2+1/2$. The numbers $\omega_k$ satisfy for $k\ge1$ the
quadratic recurrence
\begin{displaymath}
\alpha_{k-1}\omega_{k-1} +\sum_{l=0}^k \phi_l 2^{-l}\omega_{k-l}=0,
\end{displaymath}
with initial condition $\omega_0=1$, where the numbers $\phi_k$ appear in the
Airy distribution as in Definition \ref{cr:def:Airy}.
\qed
\end{fact}

\noindent \textbf{Remarks.} \textit{i)} This result has been derived using
a discrete meander, whose length and area generating function is 
described by a system of two algebraic $q$-difference equations, 
see \cite[Prop~1]{cr:N03a}. \\
\textit{ii)} We have $\mathbb E[Z]=3\sqrt{2\pi}/8$ for the mean of $Z$. 
The random variable $Z$ is uniquely determined by its moments.
The numbers $\omega_k$ appear in the asymptotic expansion \cite[Thm 3]{cr:T95}
\begin{equation*}
\Omega(s)=\frac{1-3\int_0^s\Ai(t)\,{\rm d}t}{3\Ai(s)}
\sim \sum_{k\ge 0} (-1)^k \omega_ks^{-\alpha_k}
 \qquad (s\to\infty),
\end{equation*}
where $\Ai(x)=\frac{1}{\pi}\int_0^\infty \cos(t^3/3+tx)\,{\rm d}t$ 
is the Airy function.

\medskip

Explicit expressions have been derived for the moment generating 
function and for the distribution function of $Z$.

\begin{fact}[{\cite[Thm~5]{cr:T95}}]
The moment generating function $M(t)=\mathbb E[\e^{-tZ}]$ of $Z$
satisfies the Laplace transform 
\begin{equation}\label{Ecr:meanlap}
\int_0^\infty \e^{-st} M(\sqrt{2}\,t^{3/2}) \frac{1}{t^{1/2}}\,{\rm d}t= 
\sqrt{\pi}\,\Omega(s).
\end{equation}
It is explicitly given by
\begin{equation*}
M(t)= 2^{-1/6}t^{1/3}\sum_{k=1}^\infty R_k\exp(-\beta_k 
t^{2/3}2^{-1/3})
\end{equation*}
for $\Re(t)>0$, where the numbers $-\beta_k$ are the zeroes 
of the Airy function, and where
\begin{equation*}
R_k = \frac{\beta_k(1+3\int_0^{\beta_k}\Ai(-t)
\,{\rm d}t)}{3\Ai'(-\beta_k)}.
\end{equation*}
The random variable $Z$ has a continuous density $p(y)$, with 
distribution function $R(x)=\int_0^x p(y)\,{\rm d}y$ given by
\begin{equation*}
R(x)=\frac{\sqrt{\pi}}{(18)^{1/6}x}\sum_{k=1}^\infty R_k \e^{-v_k}
v_k^{-1/3}\Ai((3v_k/2)^{2/3}),
\end{equation*}
where $v_k=(\beta_k)^3/(27x^2)$.\qed
\end{fact}

\noindent {\bf Remark.} The moment generating function and
the distribution function are obtained by two consecutive 
inverse Laplace transforms of Eq.~(\ref{Ecr:meanlap}).

\begin{theorem}
The normalised area random variables $X_m$ of 
directed convex polygons Eq.~(\ref{Ecr:norm}) satisfy
\begin{equation*}
X_m \stackrel{d}{\longrightarrow} \frac{1}{2}Z\qquad (m\to\infty),
\end{equation*}
where $Z$ is the area random variable of the Brownian 
meander as in Fact \ref{cr:def:mean}. We also have
moment convergence.
\end{theorem}

\begin{proof}

A system of $q$-difference equations for the generating 
function $Q(x,y,q)$ of directed convex polygons, counted by
width, height and area, has been given in \cite[Lemma~1.1]{cr:BF95}. 
It can be reduced to a single equation,
\begin{equation}\label{Ecr:dirconfeq}
\begin{split}
&q(qx-1)Q(x,y,q)+
\left((1+q)(P(x,y,q)+y)\right)Q(qx,y,q)+\\
&\left(xyq-y^2+P(x,y,q)(qx-y-1)\right)Q(q^2x,y,q)\\
&-q^2xy\left(y+P(x,y,q)-1\right)=0,
\end{split}
\end{equation}
where $P(x,y,q)$ is the width, height and area generating 
function of staircase polygons. Setting $q$=1 and $x=y$ 
yields the half-perimeter generating function
\begin{equation*}
g_0(x)=\frac{x^2}{\sqrt{1-4x}}.
\end{equation*}
Hence $x_c=1/4$ for the radius of convergence of $Q(x,x,1)$.

It is possible to derive from Eq.~(\ref{Ecr:dirconfeq}) a 
$q$-difference equation for the (isotropic) half-perimeter and 
area generating function $Q(x,q)=Q(x,x,q)$ of directed convex 
polygons. This is due to the symmetry $Q(x,y,q)=Q(y,x,q)$,
which results from invariance of the set of directed convex
polygons under reflection along the negative diagonal $y=-x$. 
Since this equation is quite long,
we do not give it here. 
By arguments analogous to those of the previous subsection,  
it can be deduced from this equation that all area moment generating functions $g_k(x)$ of $Q(x,1)$ are Laurent series 
in $s=\sqrt{1-4x}$, see also \cite[Prop~(4.3)]{cr:Rich05}. The leading singular exponent of $g_k(x)$, 
defined by $g_k(x)\sim h_k (1-x/x_c)^{-\alpha_k}$ as 
$x\nearrow x_c$, can be bounded from above by 
$\alpha_k\le3k/2+1/2$, see also \cite[Prop~(4.4)]{cr:Rich05} for the argument.
We apply the method of dominant
balance, in order to prove that 
$\alpha_k=3k/2+1/2$ and to yield recurrences for the coefficients 
$h_k$. We define
\begin{equation*}
\begin{split}
P(x,q)&=\frac{1}{4}+(1-q)^{1/3} F\left(\frac{1-4x}
{(1-q)^{2/3}},(1-q)^{1/3}\right),\\
Q(x,q)&=(1-q)^{-1/3} H\left(\frac{1-4x}
{(1-q)^{2/3}},(1-q)^{1/3}\right),
\end{split}
\end{equation*}
where $F(s)=F(s,0)$ has already been determined in Eq.~(\ref{Ecr:stairscal}).
We set $4x=1-s\epsilon^2$, $q=1-\epsilon^3$, and expand
the $q$-difference equation to leading order in 
$\epsilon$. We get for $H(s):=H(s,0)$ the inhomogeneous 
linear differential equation of first order
\begin{equation*}
H'(s)+4H(s)F(s)+\frac{1}{8}=0.
\end{equation*}
This implies for the coefficients $h_k$ of $H(s)=\sum_{k\ge0} h_k s^{-\alpha_k}$
and $f_k$ of $F(s)=\sum_{k\ge0}f_ks^{-\gamma_k}$ for $k\ge1$ the quadratic 
recursion
\begin{displaymath}
\alpha_{k-1}h_{k-1}+4\sum_{l=0}^k f_lh_{k-l}=0,
\end{displaymath}
where $h_0=1/16$.  Using $f_k=2^{-2k-1}\phi_k$, we obtain the 
meander recursion in Fact \ref{cr:def:mean} by setting 
$h_k=2^{-k-4}\omega_k$. It can be inferred from the functional
equation that (the analytic continuations of) all factorial moment 
generating functions are $\Delta$-regular, with $x_c=1/4$. 
Thus Lemma~\ref{Ecr:transfergx} applies, and we conclude
$X_m\stackrel{d}{\to} Z/2$.
\qed
\end{proof}

\noindent {\bf  Remarks.} \textit{i)}
The above theorem states that the limit distribution of 
area of directed convex polygons coincides, up to 
normalisation, with the area distribution of the Brownian 
meander  \cite{cr:T95}. This suggests that there might exist 
a combinatorial bijection to discrete meanders, in analogy to 
that between staircase polygons and Dyck paths. 
Up to now, a ``nice'' bijection has not been found, see 
however \cite{cr:B92,cr:DMPR01} for combinatorial bijections 
to discrete bridges.\\
\textit{ii)} The above proof relies on a $q$-difference equation
for the isotropic generating function $Q(x,x,q)$. Up to
normalisation, the meander distribution also appears
for the anisotropic model $Q(x,y,q)$, where $0<y<1/2$ is 
fixed, as can be shown by a considerably simpler calculation.
The normalisation constant coincides with that of the
isotropic model for $y=1/2$. The latter statement is also
a consequence of the fact that the height random variable
of directed polygons is asymptotically Gaussian, after
centering and normalisation. Analogous considerations apply
to the relation between isotropic and anisotropic versions
of the other polygon classes.

\subsection{Limit laws away from $(x_c,1)$}

As motivated in the introduction, limit laws in the fixed perimeter 
ensemble for $q\ne1$ are expected to be Gaussian. The same 
remark holds for the fixed area ensemble for $x\ne x_c$. 
There are partial results for the model of staircase polygons. 
The fixed area ensemble can, for $x<x_c$ and $q$ near unity, 
be analysed using Fact~\ref{cr:fact:stairsf} of the following section. 
For staircase polygons in the uniform fixed area ensemble $x=1$, the 
following result holds.

\begin{fact}[{\cite[Prop~IX.11]{cr:FS06}}]
Consider the perimeter random variable of staircase polygons
in the uniform fixed area ensemble,
\begin{equation*}
\mathbb P(\widetilde Y_n=m)=\frac{p_{m,n}}{\sum_mp_{m,n}}.
\end{equation*}
The variable $\widetilde Y_n$ has mean $\mu_n\sim \mu \cdot n$ and standard
deviation $\sigma_n\sim \sigma \sqrt{n}$, where the numbers $\mu$ and
$\sigma$ satisfy
\begin{equation*}
\mu=0.8417620156\ldots,\qquad \sigma=0.4242065326\ldots
\end{equation*}
The centred and normalised random variables
\begin{equation*}
Y_n=\frac{\widetilde Y_n-\mu_n}{\sigma_n},
\end{equation*}
converge in distribution to a Gaussian random variable.
\qed
\end{fact}

\noindent {\bf Remark.} The above result is proved using
an explicit expression for the half-perimeter and area
generating function, as a ratio of two $q$-Bessel functions.
It can be shown that this expression is meromorphic
about $(x,q)=(1,q_c)$ with a simple pole, where $q_c$ is 
the radius of convergence of the generating function 
$P(1,q)$. The explicit form of the singularity about
$(1,q_c)$ yields a Gaussian limit law.

\medskip

There are a number of results for classes of column-convex
polygons in the uniform fixed area ensemble, typically
leading to Gaussian limit laws. The upper and lower shape of 
a polygon can be described by Brownian motions. See 
\cite{cr:L96,cr:L97,cr:L99} for details.
It would be interesting to prove convergence to a Gaussian limit 
law within a more general framework, such as $q$-difference 
equations. Analogous questions for other functional equations, 
describing counting parameters such as horizontal width, 
have been studied in \cite{cr:Drmota97}.

\subsection{Self-avoiding polygons}\index{self-avoiding polygons}

A numerical analysis of self-avoiding polygons, using
data from exact enumeration \cite{cr:RGJ01,cr:RJG03}, supports the 
conjecture that the limit law of area is, up to normalisation, 
the Airy distribution.

Let $p_{m,n}$ denote the number of square lattice self-avoiding 
polygons of half-perimeter $m$ and area $n$. Exact enumeration
techniques have been applied to obtain the numbers $p_{m,n}$
for all values of $n$ for given $m\le50$. Numerical extrapolation
techniques yield very accurate estimates of the asymptotic
behaviour of the coefficients of the factorial moment
generating functions. To leading order, these are given by
\begin{equation}\label{Ecr:ampAk}
[x^m] g_k(x) =\frac{1}{k!}\sum_n (n)_k p_{m,n}\sim A_k x_c^{-m} 
m^{3k/2-3/2-1}
\qquad (m\to\infty),
\end{equation}
for positive amplitudes $A_k$. The above form has been
numerically checked \cite{cr:RGJ01,cr:RJG03} for values $k\le10$ 
and is conjectured to hold for arbitrary $k$. The value $x_c$ is the 
radius of convergence of the half-perimeter generating function of 
self-avoiding polygons. The amplitudes 
$A_k$ have been extrapolated to at least five significant 
digits. In particular, we have
\begin{displaymath}
x_c=0.14368062927(2), \qquad A_0 = 0.09940174(4),\qquad A_1=0.0397886(1),
\end{displaymath}
where the numbers in brackets denote the uncertainty in 
the last digit. An exact value of the amplitude $A_1=1/(8\pi)$ has 
been predicted \cite{cr:C94} using field-theoretic arguments.

The particular form of the exponent implies that the model
of \emph{rooted} self-avoiding polygons $\widetilde p_{m,n}=mp_{m,n}$
has the same exponents $\phi=2/3$ and $\theta=1/3$ as staircase 
polygons. In particular, it implies a square-root as dominant 
singularity of the half-perimeter generating function. 
Together with the above result for $q$-functional equations, 
this suggests that (rooted) self-avoiding polygons might obey 
the Airy distribution as a limit law of area.

A natural method to test this conjecture consists in
analysing ratios of moments, such that a normalisation
constant is eliminated. Such ratios are also called 
\emph{universal amplitude ratios}. If the conjecture were true,
we would have asymptotically
\begin{equation*}
\frac{\mathbb E[\widetilde X_m^k]}{\mathbb E[\widetilde X_m]^k}
\sim k! \frac{\Gamma(\gamma_1)^k}{\Gamma(\gamma_k)\Gamma(\gamma_0)^{k-1}} 
\frac{\phi_k\phi_0^{k-1}}{\phi_1^k}\qquad (m\to\infty),
\end{equation*}
for the area random variables $\widetilde X_m$ as in Eq.~(\ref{Ecr:xmt}).
The numbers $\phi_k$ and exponents $\gamma_k$ are those of
the Airy distribution as in Definition \ref{cr:def:Airy}.
The above form was numerically confirmed for values of $k\le10$ to
a high level of numerical accuracy. The normalisation
constant is obtained by noting that $\mathbb E[Y]=\sqrt{\pi}$.

\begin{conj}[cf \cite{cr:RGJ01,cr:RJG03}]\label{Ccr:SAPlim}
Let $p_{m,n}$ denote the number
of square lattice self-avoiding polygons of half-perimeter $m$ and area $n$.
Let $\widetilde X_m$ denote the random variable of area 
in the uniform fixed perimeter ensemble,
\begin{equation*}
\mathbb P(\widetilde X_m=n)=\frac{p_{m,n}}{\sum_n p_{m,n}}.
\end{equation*}
We conjecture that
\begin{equation*}
\frac{\widetilde X_m}{\mathbb E[\widetilde X_m]} \stackrel{d}{\longrightarrow} 
\frac{Y}{\sqrt{\pi}},
\end{equation*}
where $Y$ is Airy distributed according to Definition \ref{cr:def:Airy}.
\end{conj}

\noindent {\bf Remarks.} {\it i)} Field theoretic arguments
\cite{cr:C94} yield $A_1=1/(8\pi)$.\\
{\it ii)} References \cite{cr:RGJ01,cr:RJG03} contain
conjectures for the scaling function of self-avoiding polygons and rooted
self-avoiding polygons, see the following section. In fact, the numerical analysis
in \cite{cr:RGJ01,cr:RJG03} mainly concerns the area amplitudes 
$A_k$, which determine the limit distribution of area.\\
{\it iii)}
The area law of self-avoiding polygons has also been 
studied \cite{cr:RGJ01,cr:RJG03} on the triangular and 
hexagonal lattices. As for the square lattice, the 
area limit law appears to be the Airy distribution, up
to normalisation.\\
{\it iv)} It is an open question whether there are
non-trivial counting parameters other than the area, whose limit
law (in the fixed perimeter ensembles) coincides between 
self-avoiding polygons and staircase polygons. See 
\cite{cr:Rich06} for a negative 
example. This indicates that underlying stochastic
processes must be quite different.\\
{\it v)} A proof of the above conjecture is an outstanding open 
problem. It would be interesting to analyse the emergence of 
the Airy distribution using stochastic Loewner evolution \cite{cr:LSW02}.
Self-avoiding polygons at criticality are conjectured
to describe the hull of critical percolation clusters
and the outer boundary of two-dimensional Brownian 
motion \cite{cr:LSW02}.

\medskip

A numerical analysis of the fixed area ensemble along the
above lines again shows behaviour similar to that of staircase polygons.
This supports the following conjecture.

\begin{conj}
Consider the perimeter random variable of self-avoiding polygons
in the uniform fixed area ensemble,
\begin{equation*}
\mathbb P(\widetilde Y_n=m)=\frac{p_{m,n}}{\sum_mp_{m,n}}.
\end{equation*}
The random variable $\widetilde Y_n$ is conjectured to have mean 
$\mu_n\sim \mu \cdot n$ and standard deviation $\sigma_n\sim \sigma 
\sqrt{n}$, where the numbers $\mu$ and $\sigma$ satisfy
\begin{equation*}
\mu=1.855217(1),\qquad \sigma^2=0.3259(1),
\end{equation*}
where the number in brackets denotes the uncertainty in the last digit.
The centred and normalised random variables
\begin{equation*}
Y_n=\frac{\widetilde Y_n-\mu_n}{\sigma_n},
\end{equation*}
are conjectured to converge in distribution to a Gaussian random variable.
\end{conj}

\medskip

The above conjectures, together with the results of the previous
subsection, also raise the question whether rooted square-lattice 
self-avoiding polygons, counted by half-perimeter and area, might 
satisfy a $q$-functional equation. In particular, 
it would be interesting to consider whether rooted self-avoiding 
polygons might satisfy
\begin{equation}\label{Ecr:feqnconj}
P(x)=G(x,P(x)),
\end{equation}
for some power series $G(x,y)$ in $x,y$. If the perimeter
generating function $P(x)$ is not algebraic, this excludes
polynomials $G(x,y)$ in $x$ and $y$. Note that the anisotropic 
perimeter generating function of self-avoiding polygons is 
not $D$-finite \cite{cr:Rech06}. 
It is thus unlikely that the isotropic perimeter generating 
function is $D$-finite and, in particular, algebraic.
On the other hand, solutions of Eq.~(\ref{Ecr:feqnconj}) 
need not be algebraic nor $D$-finite. An example is the Cayley 
tree generating function $T(x)$ satisfying $T(x)=x\exp(T(x))$, 
see \cite{cr:FGS05}.

\subsection{Punctured polygons}\index{punctured polygons}

Punctured polygons are self-avoiding polygons with internal holes,
which are also self-avoiding polygons. The polygons are also 
mutually avoiding. The perimeter of a punctured polygon is the
sum of the lengths of its boundary curves, the area of a punctured polygon
is the area of the outer polygon minus the area of
the holes. Apart from intrinsic combinatorial interest, models 
of punctured polygons may be viewed as arising from 
two-dimensional sections of three-dimensional self-avoiding 
vesicles. Counted by area, they may serve as an
approximation to the polyomino model. 

\smallskip

We consider, for a given subclass of self-avoiding polygons, punctured
polygons with holes from the same subclass. The case
of a bounded number of punctures of bounded size can be
analysed in some generality. The case of a bounded
number of punctures of unbounded size leads to simple
results if the critical perimeter generating function
of the model without punctures is finite.

\smallskip

For a given subclass of self-avoiding polygons, the number 
$p_{m,n}$ denotes the number of polygons with half-perimeter 
$m$ and area $n$. Let $p_{m,n}^{(r,s)}$ denote the number of 
polygons with $r\ge1$ punctures whose half-perimeter sum 
equals $s$. Let $p_{m,n}^{(r)}$ denote the number of polygons with 
$r\ge1$ punctures of arbitrary size.

\begin{theorem}[{\cite[Thms~1,2]{cr:RJG06}}]
Assume that, for a class of self-avoiding polygons without 
punctures, the area moment coefficients $p_m^{(k)}=
\sum_{n\ge0} n^k p_{m,n}$ have, for $k\in\mathbb N_0$, the 
asymptotic form
\begin{equation*}
p_m^{(k)} \sim A_k x_c^{-m} m^{\gamma_k-1} \qquad 
(m\to\infty),
\end{equation*} 
for numbers $A_k>0$, for $0<x_c\le 1$ and for $\gamma_k=
(k-\theta)/\phi$, where $0<\phi<1$.
Let $g_0(x)=\sum_{m\ge0} p_m^{(0)}x^m$ denote the half-perimeter 
generating function. 

Then, the area moment coefficient $p_m^{(r,k,s)}=\sum_n n^k 
p_{m,n}^{(r,s)}$ of the polygon class with $r\ge1$ punctures 
whose half-perimeter sum equals $s$ is, for $k\in\mathbb N_0$,
asymptotically given by 
\begin{equation*}
p_m^{(r,k,s)} \sim A_k^{(r,s)} x_c^{-m}m^{\gamma_{k+r}-1}
\qquad (m\to\infty),
\end{equation*}
where $A_{k}^{(r,s)}=\frac{A_{k+r}}{r!} x_c^s[x^s](g_0(x))^r$.

If $\theta>0$, the area moment coefficient $p_m^{(r,k)}=\sum_n n^k 
p_{m,n}^{(r)}$ of the polygon class with $r\ge1$ 
punctures of arbitrary size satisfies, for $k\in\mathbb N_0$, asymptotically
\begin{equation*}
p_m^{(r,k)}\sim
A_k^{(r)}x_c^{-m} m^{\gamma_{k+r}-1}
\qquad (m\to\infty),
\end{equation*}
where the amplitudes $A_k^{(r)}$ are given by
\begin{equation*}
A_k^{(r)}=\frac{A_{k+r}(g_0(x_c))^r}{r!}.
\end{equation*}
\qed
\end{theorem}

\noindent {\bf  Remarks.} {\it i)}
The basic argument in the proof of the preceding result involves
an estimate of interactions of hole polygons with one another
or with the boundary of the external polygon, which are shown to
be asymptotically irrelevant. This argument also applies
in higher dimensions, as long as the exponent $\phi$ satisfies 
$0<\phi<1$.  \\
{\it ii)} In the case of an infinite critical perimeter generating function, 
such as for subclasses of convex polygons, 
boundary effects are asymptotically relevant, if punctures 
of unbounded size are considered. The case of an unbounded number 
of punctures, which approximates the polyomino problem, is unsolved.\\
{\it iii)} The above result leads to new area limit 
distributions. For rectangles with $r$ punctures of bounded size,
we get $\beta_{r+1,1/2}$ as the limit distribution of area. For 
staircase polygons with punctures, we obtain generalisations of 
the Airy distribution, which are discussed in \cite{cr:RJG06}. In contrast, 
for Ferrers diagrams with punctures of bounded size, the limit
distribution of area stays concentrated.\\
{\it iv)} The theorem also applies to models of punctured
polygons, which do not satisfy an algebraic $q$-difference equation.
An example is given by staircase polygons with a staircase hole
of unbounded size, whose perimeter generating function is 
not algebraic \cite{cr:GJ06b}.

\subsection{Models in three dimensions}\label{cr:sec3d}

There are very few results for models in higher dimensions, notably 
for models on the cubic lattice. There
are a number of natural counting parameters for such objects. We
restrict consideration to area and volume, which is the 
three-dimensional analogue of perimeter and area of two-dimensional models. 

One prominent model is self-avoiding surfaces on the cubic lattice,
also studied as a model of three-dimensional vesicle collapse. 
We follow the review in \cite{cr:V98} (see also the references therein) 
and consider closed orientable surfaces of genus zero, i.e., surfaces 
homeomorphic to a sphere. Numerical studies indicate that the 
surface generating function displays a square-root $\gamma=-1/2$ 
as the dominant singularity.

Consider the fixed surface area ensemble with weights proportional to 
$q^n$, with $n$ the volume of the surface. One expects a deflated phase 
(branched polymer phase)
for small values of $q$ and an inflated phase (spherical phase) for large 
values of $q$. In the deflated phase, the mean volume of a surface 
should grow proportionally to the area $m$ of the surface, in the inflated
phase the mean volume should grow like $m^{3/2}$ with the surface.
Numerical simulations suggest a phase transition at $q=1$ with
exponent $\phi=1$. This indicates that a typical surface resembles
a branched polymer, and a concentrated distribution of volume
is expected. Note that this behaviour differs from that of the 
two-dimensional model of self-avoiding polygons.

Even relatively simple subclasses of self-avoiding surfaces such as rectangular
boxes \cite{cr:MJ05} and plane partition vesicles \cite{cr:JM06},
generalising the two-dimensional models of rectangles and Ferrers diagrams,
display complicated behaviour. Let $p_{m,n}$ denote the
number of surfaces of area $m$ and volume $n$ and consider the
generating function $P(x,q)=\sum_{m,n}p_{m,n}x^mq^n$. For rectangular
box vesicles, we apparently have $P(x,1)\sim A|\log(1-x)|/(1-x)^{3/2}$ as $x\to1^-$,
some some constant $A>0$, see \cite[Eq~(35)]{cr:MJ05}.
In the fixed surface area ensemble, a linear polymer phase $0<q<1$
is separated from a cubic phase $q>1$. At $q=1$, we have $\phi=2/3$, 
such that typical rectangular boxes are expected to attain a cubic shape.
We expect a limit distribution which is concentrated. For plane
partition vesicles, it is conjectured on the basis of numerical simulations  
\cite[Sec~4.1.1]{cr:JM06} that $P(x,1)\sim A\exp(\alpha/(x_c-x)^{1/3})
/(x_c-x)^\gamma$, where $\gamma\approx1.7$ at $x_c=0.8467(3)$, for
non-vanishing constants $A$ and $\alpha$. It is expected that $\phi=1/2$.

As in the previous subsection, three-dimensional models of punctured
vesicles may be considered. The above arguments hold, if the exponent
$\phi$ satisfies $0<\phi<1$. A corresponding result for punctures
of unbounded size can be stated if the critical surface area generating
function is finite.

\subsection{Summary}

In this section, we described methods to extract asymptotic 
area laws for polygon models on the square lattice, and we 
applied these to various classes of polygons. Some of the 
laws were found to coincide with those of the (absolute) area 
under a Brownian excursion and a Brownian meander. A 
combinatorial explanation for the latter result has not been given.
Is there a simple polygon model with the same area limit law as the area 
under a Brownian bridge? The connection to stochastics 
deserves further investigation. In particular, it would be interesting 
to identify underlying stochastic processes. For an approach to 
a number of different random combinatorial structures starting 
from a probabilistic viewpoint, see \cite{cr:P06}.

Area laws of polygon models in the uniform fixed perimeter
ensemble $q=1$ have been understood in some generality, 
by an analysis of the singular behaviour of $q$-functional 
equations about the point $(x,q)=(x_c,1)$. Essentially,
the type of singularity of the half-perimeter generating
function determines the limit law. A refined analysis can 
be done, leading to local limit laws and providing 
convergence rates. Also, limit distributions
describing corrections to the asymptotic behaviour
can be derived. They seem to coincide with distributions
arising in models of punctured polygons, see \cite{cr:RJG06}.

For non-uniform ensembles, concentrated distributions are 
expected, but general results, e.g. for $q$-functional equations,
are lacking. These may be obtained by multivariate singularity 
analysis, see also \cite{cr:Drmota97,cr:Lladser03}.

The underlying structure of $q$-functional equations appears
in a number of other combinatorial models, such as models
of two-dimensional directed walks, counted by length and area 
between the walk and the $x$-axis, models of simply generated trees, counted
by the number of nodes and path length, and models which
appear in the average case analysis of algorithms, see 
\cite{cr:FL01,cr:FS06}. Thus, the above methods and results can be applied
to such models. In statistical physics, this mainly
concerns models of (interacting) directed walks, see 
\cite{cr:Janse05} for a review. There is also an approach 
to the behaviour of such walks from a stochastic viewpoint, 
see e.g.~the review \cite{cr:HK01}.

There are exactly solvable polygon models, which do not
satisfy an algebraic $q$-difference equation, such as three-choice
polygons \cite{cr:GJ06a}, punctured staircase polygons \cite{cr:GJ06b}, 
prudent polygon subclasses \cite{cr:S08},
and possibly diagonally convex polygons. For a rigorous
analysis of the above models, it may be necessary
to understand $q$-difference equations with more
general holonomic solutions, as $q$ approaches unity.

Focussing on self-avoiding polygons, it might be
interesting to analyse whether the perimeter generating
function of rooted self-avoiding polygons might satisfy an implicit
equation Eq.~(\ref{Ecr:feqnconj}). Asymptotic 
properties of the area can possibly be studied using stochastic
Loewner evolution \cite{cr:LSW02}. Another open question
concerns  the area limit law for $q\ne1$ or the perimeter 
limit law for $x\ne x_c$, where Gaussian behaviour is 
expected. At present, even the simpler question of analyticity 
of the critical curve $x_c(q)$ for $0<q<1$ is open.

\medskip

Most results of this section concerned area limit
laws of polygon models. Similarly, one can ask for
perimeter laws in the fixed area ensemble. Results
have been given for the uniform ensemble. Generally,
Gaussian limit laws are expected away from criticality, 
i.e., away from $x=x_c$.
Perimeter laws are more difficult to extract from a 
$q$-functional equation than area laws. We will 
however see in the following section that, surprisingly, 
under certain conditions, knowledge of the area limit 
law can be used to infer the perimeter limit law at criticality.

\section{Scaling functions}

From a technical perspective, the focus in the previous section was 
on the singular behaviour of the single-variable factorial moment generating 
function $g_k(x)$ Eq.~(\ref{Ecr:gengk}), and on the associated 
asymptotic behaviour of their coefficients. This yielded the limiting 
area distribution of some polygon models. 

In this section, we discuss the more 
general problem of the singular behaviour of the 
two-variable perimeter and area generating function 
of a polygon model. Near the special point $(x,q)=(x_c,1)$, 
the perimeter and area generating function 
$P(x,q)=\sum_{m\ge0} p_m(q)x^m=\sum_{n\ge0} a_n(x) q^n$ is expected 
to be approximated by a scaling function, and the 
corresponding coefficient functions $p_m(q)$ and $a_n(x)$ are 
expected to be approximated by finite
size scaling functions. As we will see, scaling functions
encapsulate information about the limit distributions discussed in the
previous section, and thus have a probabilistic interpretation.

\smallskip

We will give a focussed review, guided by exactly solvable 
examples, since singularity analysis of multivariate 
generating functions is, in contrast to the one-variable case,
not very well developed, see \cite{cr:PW06} for a recent overview. 
Methods of particular interest to polygon models concern 
asymptotic expansions about multicritical points, which are 
discussed for special examples in \cite{cr:O74,cr:BH86}. Conjectures 
for the behaviour of polygon models about multicritical points 
arise from the physical theory of tricritical scaling \cite{cr:G73}, 
see the review \cite{cr:LS84}, which has been adapted to
polygon models \cite{cr:BOP93,cr:BO95}. There are few rigorous results
about scaling behaviour of polygon models, which we will
discuss.  This will complement the exposition in \cite{cr:J00}.
See also \cite[Ch 9]{cr:Grimmett99} for the related subject of 
scaling in percolation.

\subsection{Scaling and finite size scaling}

The half-perimeter and area generating function of
a polygon model $P(x,q)$ about $(x,q)=(x_c,1)$ is 
expected to be approximated by a scaling function. This 
is motivated by the following heuristic 
argument.
Assume that the factorial area moment generating functions $g_k(x)$ 
Eq.~(\ref{Ecr:gengk}) have, for values $x<x_c$, a local expansion 
about $x=x_c$ of the form
\begin{equation*}
g_k(x)=\sum_{l\ge0} \frac{f_{k,l}}{(1-x/x_c)^{\gamma_{k,l}}},
\end{equation*}
where $\gamma_{k,l}=(k-\theta_l)/\phi$ and $\theta_{l+1}>\theta_l$. 
Disregarding questions of analyticity, we argue
\begin{displaymath}
\begin{split}
P(x,q) &\approx \sum_{k\ge0} (-1)^k\left(\sum_{l\ge0}
\frac{f_{k,l}}{(1-x/x_c)^{\gamma_{k,l}}}\right)(1-q)^k\\
&\approx \sum_{l\ge0}(1-q)^{\theta_l}\left(\sum_{k\ge0}
(-1)^kf_{k,l} \left( \frac{1-x/x_c}{(1-q)^{\phi}}\right)^{-\gamma_{k,l}}
\right).
\end{split}
\end{displaymath}
In the above calculation, we replaced $P(x,q)$ by its Taylor series
about $q=1$, and then replaced the Taylor coefficients by their
expansion about $x=x_c$. The preceding heuristic calculation has, 
for some polygon models and on a formal level, a rigorous counterpart, 
see the previous section.
In the above expression, the r.h.s. depends on series 
${\cal F}_l(s)=\sum_{k\ge0}(-1)^kf_{k,l} s^{-\gamma_{k,l}}$ of 
a single variable of combined argument $s=(1-x/x_c)/(1-q)^\phi$.
Restricting to the leading term $l=0$, this motivates the following 
definition. For $\phi>0$ and $x_c>0$, we define for numbers 
$s_{-},s_{+}\in[-\infty,+\infty]$ the domain
\begin{equation*}
D(s_{-},s_{+})=\{(x,q)\in (0,\infty)\times(0,1): s_{-}<(1-x/x_c)/(1-q)^\phi<s_{+})\}.
\end{equation*}

\begin{definition}[Scaling function]\label{cr:def:sf}\index{scaling function}
For numbers $p_{m,n}$ with generating function 
$P(x,q)=\sum_{m,n}p_{m,n}x^mq^n$, let Assumption \ref{Acr:pol} be satisfied. 
Let $0<x_c\le1$ be the radius of convergence 
of $P(x,1)$. Assume that there exist constants $s_{-}, s_{+}\in[-\infty,+\infty]$
satisfying $s_{-}<s_{+}$ and a function ${\cal F}:(s_{-},s_{+})\to\mathbb R$, such 
that $P(x,q)$ satisfies, for real constants $\theta$ and $\phi>0$,
\begin{equation}\label{Ecr:defscal}
P^{(sing)}(x,q) \sim (1-q)^\theta {\cal F}\left(\frac{1-x/x_c}{(1-q)^{\phi}}\right)
\qquad (x,q)\to (x_c,1) \mbox{ in } D(s_{-},s_{+}).
\end{equation}
Then, the function ${\cal F}(s)$ is called an \emph{(area) scaling 
function}, and $\theta$ and $\phi$ are called \emph{critical exponents}.
\end{definition}

\noindent {\bf  Remarks.} 
\textit{i)}
In analogy to the one-variable case, the above asymptotic equality
means that there exists a power series $P^{(reg)}(x,q)$ convergent for 
$|x|<x_1$ and $|q|<q_1$, where $x_1>x_c$ and $q_1>1$, such
that the function $P^{(sing)}(x,q):=P(x,q)-P^{(reg)}(x,q)$ is 
asymptotically equal to the r.h.s..\\
\textit{ii)}
Due to the region $D(s_{-},s_{+})$ where the limit $(x,q)\to(x_c,1)$ is 
taken, admissible values $(x,q)$ satisfy $0<q<1$ and $0<x<x_0(q)$, 
where $x_0(q)= x_c(1-s_{-}(1-q)^\phi)$, if $s_{-}\ne-\infty$. Thus, in this 
case, the critical curve $x_c(q)$ satisfies $x_c(q)\ge x_0(q)$ as $q$ 
approaches unity. Note that equality need not hold in general.
\\
\textit{iii)} The method of dominant balance was originally 
applied in order to obtain a defining equation for a 
scaling function ${\cal F}(s)$ from a given functional 
equation of a polygon model. This assumes the existence 
of a scaling function, together with additional analyticity 
properties. See \cite{cr:PB95,cr:RGJ01,cr:Rich02}.\\
\textit{iv)} For particular examples, an analytic scaling function 
${\cal F}(s)$ exists, with an asymptotic expansion about infinity,
and the area amplitude series $F(s)$ agrees with the asymptotic series, 
see below.\\
\textit{v)}
There is an alternative definition of a scaling function \cite{cr:FGW91}
by demanding 
\begin{equation}\label{Ecr:alt}
P^{(sing)}(x,q) \sim \frac{1}{(1-x/x_c)^{-\theta/\phi}}
{\cal H}\left(\frac{1-q}{(1-x/x_c)^{1/\phi}}\right)
\qquad (x,q)\to (x_c,1)
\end{equation}
in a suited domain, for a function ${\cal H}(t)$ of argument 
$t=(1-q)/(1-x/x_c)^{1/\phi}$. Such a scaling form is also 
motivated by the above argument. One may then call such a function 
${\cal H}(t)$ a \emph{perimeter scaling function}. If ${\cal F}(s)$ is a scaling 
function, then a function ${\cal H}(t)$, satisfying 
Eq.~(\ref{Ecr:alt}) in a suited domain, is given by
\begin{equation*}
{\cal H}(t)=t^\theta {\cal F}(t^{-\phi}).
\end{equation*}

\medskip

If $s_{-}\le0$ and $s_{+}=\infty$, the particular scaling 
form Eq.~(\ref{Ecr:defscal}) implies a certain asymptotic 
behaviour of the critical area generating function and of 
the half-perimeter generating function. The following lemma is
a consequence of Definition~\ref{cr:def:sf}.

\begin{lemma}\label{cr:lem:scalfct}
Let the assumptions of Definition \ref{cr:def:sf} be satisfied.

\begin{itemize}

\item[\textit{i)}]
If $s_{+}=\infty$ and if the scaling function ${\cal F}(s)$ has 
the asymptotic behaviour
\begin{equation*}
{\cal F}(s)\sim f_0 s^{-\gamma_0} \qquad (s\to\infty),
\end{equation*}
then $\gamma_0=-\frac{\theta}{\phi}$, and the half-perimeter 
generating function $P(x,1)$ satisfies
\begin{displaymath}
P^{(sing)}(x,1)\sim f_0(1-x/x_c)^{\theta/\phi} \qquad (x\nearrow x_c).
\end{displaymath}

\item[\textit{ii)}]
If $s_{-}\le0$ and if the scaling function $\mathcal{F}(s)$ has the asymptotic behaviour
\begin{equation*}
{\cal F}(s)\sim h_0 s^{\alpha_0} \qquad (s\searrow0),
\end{equation*}
then $\alpha_0=0$, and the critical area generating function 
$P(x_c,q)$ satisfies
\begin{displaymath}
P^{(sing)}(x_c,q)\sim h_0(1-q)^\theta \qquad (q\nearrow 1).
\end{displaymath}

\end{itemize}
\qed
\end{lemma}

A sufficient condition for equality of the area amplitude
series and the scaling function is stated in the following lemma,
which is an extension of Lemma \ref{cr:lem:scalfct}.

\begin{lemma}\label{cr:lem:l2}
Let the assumptions of Definition \ref{cr:def:sf} be satisfied.

\begin{itemize}

\item[\textit{i)}]
Assume that the relation Eq.~(\ref{Ecr:defscal}) 
remains valid under arbitrary differentiation w.r.t. $q$. 
If $s_{+}=\infty$, if the scaling function ${\cal F}(s)$ has an 
asymptotic expansion
\begin{equation*}
{\cal F}(s)\sim \sum_{k\ge0}(-1)^kf_k s^{-\gamma_k}\qquad (s\to\infty),
\end{equation*}
and if an according asymptotic expansion is true for arbitrary
derivatives, then the following statements hold.

\begin{itemize}
\item[a)]
The exponent $\gamma_k$ is, for $k\in\mathbb N_0$, given by
\begin{displaymath}
\gamma_k=\frac{k-\theta}{\phi}.
\end{displaymath}
\item[b)]
The scaling 
function ${\cal F}(s)$ determines the asymptotic behaviour of the 
factorial area moment generating functions via
\begin{equation*}
\left(\frac{1}{k!}\left.\frac{\partial^k}{\partial q^k}P(x,q)
\right|_{q=1}\right)^{(sing)}
\sim \frac{f_k}{(1-x/x_c)^{\gamma_k}} \qquad (x\nearrow x_c).
\end{equation*}
\end{itemize}

\item[\textit{ii)}]
Assume that the relation Eq.~(\ref{Ecr:defscal}) 
remains valid under arbitrary differentiation w.r. to $x$.
If $s_{-}\le0$, and if the scaling function $\mathcal{F}(s)$ has 
an asymptotic expansion
\begin{equation*}
\mathcal{F}(s)\sim\sum_{k\ge0}(-1)^kh_k s^{\alpha_k} \qquad (s\searrow0),
\end{equation*}
and if an according asymptotic expansion is true for arbitrary
derivatives, then the following statements hold.

\begin{itemize}
\item[a)]
The exponent $\alpha_k$ is, for $k\in\mathbb N_0$, given by $\alpha_k=k$.
\item[b)]
The scaling function determines 
the asymptotic behaviour of the factorial perimeter moment generating 
functions at $x=x_c$ via
\begin{equation*}
\begin{split}
\left(\frac{1}{k!}\left.\frac{\partial^k}{\partial x^k}P(x,q)
\right|_{x=x_c}\right)^{(sing)}
\sim \frac{x_c^{-k}h_k}{(1-q)^{\beta_k}} \qquad (q\nearrow 1),\\
\end{split}
\end{equation*}
where $\beta_k=k\phi-\theta$.

\end{itemize}

\end{itemize}
\qed
\end{lemma}

\noindent {\bf Remarks.} 
Lemma \ref{cr:lem:l2} states conditions under which
the area amplitude series coincides with the scaling function.
Given these conditions, the scaling function also determines the
perimeter law of the polygon model at criticality.

\medskip 

In the one-variable case, the singular behaviour of
a generating function translates, under suitable assumptions,
to the asymptotic behaviour of its coefficients. We
sketch the analogous situation for the asymptotic
behaviour of a generating function involving a scaling function.

\begin{definition}[Finite size scaling function]\index{finite size scaling function}
For numbers $p_{m,n}$ with generating function 
$P(x,q)=\sum_{m,n}p_{m,n}x^mq^n$, let Assumption \ref{Acr:pol} be 
satisfied. Let $0<x_c\le1$ be the radius of convergence 
of the generating function $P(x,1)$.

\begin{itemize}

\item[\textit{i)}]
Assume that there exist a number $t_{+}\in(0,\infty]$ and a
function $f:[0,t_{+}]\to\mathbb R$, such that the perimeter coefficient
function asymptotically satisfies, for real constants $\gamma_0$ and $\phi>0$,
\begin{equation*}
[x^m]P(x,q) \sim x_c^{-m}m^{\gamma_0-1} 
f(m^{1/\phi}(1-q)) \qquad (q,m)\to(1,\infty),
\end{equation*}
where the limit is taken for $m$ a positive integer and for real $q$, such 
that $m^{1/\phi}(1-q)\in[0,t_{+}]$. Then, the function $f(t)$ is called a 
\emph{finite size (perimeter) scaling function}.

\item[\textit{ii)}]
Assume that there exist constants $t_{-}\in[-\infty,0)$, $t_{+}\in(0,\infty]$, 
and a function $h:[t_{-},t_{+}]\to\mathbb R$, such that
the area coefficient function asymptotically satisfies, for real constants 
$\beta_0$ and $\phi>0$,
\begin{equation*}
[q^n]P(x,q) \sim n^{\beta_0-1} 
h(n^\phi(1-x/x_c)) \qquad (x,n)\to(x_c,\infty),
\end{equation*}
where the limit is taken for $n$ a positive integer and real $x$, such 
that $n^\phi(1-x/x_c)\in[t_{-},t_{+}]$. Then, the function $h(t)$ is called a 
\emph{finite size (area) scaling function}.

\end{itemize}
\end{definition}

\noindent \textbf{Remarks.} \textit{i)} The following heuristic calculation 
motivates the expectation that a finite size scaling function approximates
the coefficient function. For the perimeter coefficient function, 
assume that the exponents $\gamma_k$ of the factorial area moment generating 
functions are of the special form $\gamma_k=(k-\theta)/\phi$. We argue
\begin{displaymath}
\begin{split}
[x^m]P(x,q) &\approx [x^m] \sum_{k=0}^\infty (-1)^k
\frac{f_k}{(1-x/x_c)^{\gamma_k}}(1-q)^k\\
&\approx x_c^{-m}m^{\gamma_0-1}\sum_{k=0}^\infty (-1)^k
\frac{f_k}{\Gamma(\gamma_k)}\left( m^{1/\phi}(1-q)\right)^k.
\end{split}
\end{displaymath}
In the above expression, the r.h.s. depends on a function $f(t)$ of a single 
variable of combined argument $t=m^{1/\phi}(1-q)$.

For the area coefficient function, we assume that 
$\beta_k=k\phi-\theta$ and argue as above,
\begin{displaymath}
\begin{split}
[q^n]P(x,q)&\approx [q^n] \sum_{k=0}^\infty (-1)^k
\frac{h_k}{(1-q)^{\beta_k}}(1-x/x_c)^k\\
&\approx n^{\beta_0-1}\sum_{k=0}^\infty (-1)^k
\frac{h_k}{\Gamma(\beta_k)}\left( n^{\phi}(1-x/x_c)\right)^k.
\end{split}
\end{displaymath}
In the above expression, the r.h.s. depends on a function $h(t)$ of a single 
variable of combined argument $t=n^{\phi}(1-x/x_c)$.\\
\textit{ii)} The above argument suggests that a scaling function 
and a finite size scaling function may be related by a Laplace 
transformation. A comparison with Eq.÷(\ref{Ecr:wat}) leads one to 
expect that finite size scaling functions are moment generating 
functions of the limit laws of area and perimeter.
\\
\textit{iii)} Sufficient conditions under which knowledge of a scaling function
implies the existence of a finite size scaling function have been given 
for the finite size area scaling function \cite{cr:BO95} using
Darboux's theorem.

\medskip

A scaling function describes the leading singular behaviour
of the generating function $P(x,q)$ in some region about 
$(x,q)=(x_c,1)$. A particular form of subsequent correction 
terms has been argued for at the beginning of the section.

\begin{definition}[Correction-to-scaling functions]\index{correction-to-scaling function}
For numbers $p_{m,n}$ with generating function 
$P(x,q)=\sum_{m,n}p_{m,n}x^mq^n$, let Assumption \ref{Acr:pol} 
be satisfied. 
Let $0<x_c\le1$ be the radius of convergence 
of the generating function $P(x,1)$. 
Assume that there exist constants $s_{-},s_{+}\in[-\infty,+\infty]$ 
satisfying $s_{-}<s_{+}$, and functions ${\cal F}_l:(s_{-},s_{+})
\to\mathbb R$ for $l\in\mathbb N_0$, such that the generating 
function $P(x,q)$ satisfies, for real constants $\phi>0$ and $\theta_l$, 
where $\theta_{l+1}>\theta_l$,
\begin{equation*}
P^{(sing)}(x,q) \sim \sum_{l\ge0}(1-q)^{\theta_l} 
{\cal F}_l\left(\frac{1-x/x_c}{(1-q)^{\phi}}\right)
\qquad (x,q)\to (x_c,1) \mbox{ in } D(s_{-},s_{+}).
\end{equation*}
Then, the function ${\cal F}_0(s)$ is a scaling function, and 
for $l\le1$, the functions ${\cal F}_l(s)$ are called
\emph{correction-to-scaling functions}.
\end{definition}

\noindent {\bf Remarks.} \textit{i)} In the above context, the symbol $\sim$ 
denotes a (generalised) asymptotic expansion (see also \cite[Ch~1]{cr:O74}): 
Let $(G_k(\boldsymbol{x}))_{k\in\mathbb N_0}$ be a sequence of
(multivariate) functions satisfying for all $k$ the estimate $G_{k+1}(\boldsymbol{x})=o
(G_k(\boldsymbol{x}))$ as $\boldsymbol{x}\to\boldsymbol{x}_c$ 
in some prescribed region. For a function $G(\boldsymbol{x})$, we then write
$G(\boldsymbol{x})\sim \sum_{k=0}^{\infty} G_k(\boldsymbol{x})$
as $\boldsymbol{x}\to\boldsymbol{x}_c$, if for all $n$ we have $G(\boldsymbol{x})= 
\sum_{k=0}^{n-1} G_k(\boldsymbol{x})+\mathcal{O}(G_n(\boldsymbol{x}))$
as $\boldsymbol{x}\to\boldsymbol{x}_c$. \\
\textit{ii)} The previous section yielded effective methods for obtaining area 
amplitude functions. These are candidates for correction-to-scaling functions,
see also \cite{cr:Rich02}.

\subsection{Squares and rectangles}\index{squares}\index{rectangles}

We consider the models of squares and rectangles, whose scaling 
behaviour can be explicitly computed. Their half-perimeter and area
generating function can be written as a single sum, to which the 
Euler-MacLaurin summation formula \cite[Ch~8]{cr:O74} can be applied.
We first discuss squares. 

\begin{fact}[{cf \cite[Thm~2.4]{cr:J04}}]
For $0<x,q<1$, the generating function 
$P(x,q)=\sum_{m=0}^\infty x^mq^{m^2/4}$ of squares,
counted by half-perimeter and area, is given by
\begin{displaymath}
P(x,q) = \frac{1}{\sqrt{|\log q|}}{\cal F}\left(\frac{|\log x|}
{\sqrt{|\log q|}}\right)+\frac{1}{2}+R(x,q),
\end{displaymath}
with ${\cal F}(s)=\sqrt{\pi}\e^{s^2}\erfc(s)$, where the 
remainder term $R(x,q)$ is bounded by
\begin{displaymath}
|R(x,q)| \le \frac{1}{6}|\log x|.
\end{displaymath}
\qed
\end{fact}

\noindent \textbf{Remarks.} \textit{i)} The remainder term differs from that 
in \cite[Thm~2.4]{cr:J04}, where it was estimated by an integral with lower 
bound one instead of zero \cite[Eq.~(46)]{cr:J04}.\\
\textit{ii)} With $x_c=1$, $s_{-}=0$ and $s_{+}=\infty$, the function ${\cal F}(s)$ is a scaling function
according to the above definition. The remainder term is uniformly 
bounded in any rectangle $[x_0,1)\times[q_0,1)$ for $0<x_0,q_0<1$, and so 
the approximation is uniform in this rectangle.\\ 
\textit{iii)} The generating function $P(x,q)$ satisfies the quadratic $q$-difference equation
$P(x,q)=1+xq^{1/4}P(q^{1/2}x,q)$.  Using the methods of the previous section,
the area amplitude series of the model can be derived. It coincides
with the above scaling function ${\cal F}(s)$. This particular form is expected, since
the distribution of area is concentrated, $p(x)=\delta(x-1/4)$, compare 
also with Ferrers diagrams.\\
\textit{iv)} It has not been studied whether the scaling region can be extended to
values $x>1$ near $(x,q)=(1,1)$. It can be checked that the scaling function 
${\cal F}(s)$ also determines the asymptotic behaviour of the perimeter moment 
generating functions, via its expansion about the origin. As expected, they 
indicate a concentrated distribution.

\medskip

The half-perimeter and area generating function of rectangles
is given by
\begin{displaymath}
P(x,q)=\sum_{r=1}^\infty \sum_{s=1}^\infty x^{r+s}q^{rs}
=\sum_{r=1}^\infty \frac{x(qx)^r}{1-q^rx}.
\end{displaymath}
We have $P(x,1)=x^2/(1-x)^2$, and it can be shown that
$P(1,q)\sim -\frac{\log(1-q)}{1-q}$
as $q\nearrow1$, see \cite{cr:PO95,cr:J04}. The latter result implies
that a scaling form as in Definition \ref{cr:def:sf}, with 
$s_-\le0$, does not exist for rectangles. We have the following
result.

\begin{fact}[{\cite[Thm~3.4]{cr:J04}}]
For $0<q<1$ and $0<qx<1$, the generating 
function $P(x,q)$ of rectangles satisfies
\begin{displaymath}
P(x,q) = \frac{x}{|\log q|} \left( \frac{|\log q|}{|\log x|}
-\LerchPhi\left(qx,1,\frac{|\log x|}{|\log q|}\right)\right)
+R(x,q),
\end{displaymath}
with the Lerch Phi-function $\LerchPhi(z,a,v)=\sum_{n=0}^\infty 
\frac{z^n}{(v+n)^a}$, where the remainder term $R(x,q)$ is bounded by
\begin{displaymath}
|R(x,q)| \le \frac{x^2q}{1-qx}\left( \frac{1}{2}+\frac{|\log x|}{6}\right)
+ \frac{x^2q}{(1-qx)^2}\frac{|\log q|}{6}.
\end{displaymath}
\qed
\end{fact}

\noindent \textbf{Remarks.} \textit{i)} The theorem implies that, for every 
$q_0\in(0,1)$, the function $(1-qx)^2P(x,q)$ is uniformly approximated 
for points $(x,q)$ satisfying $q_0<q<1$ and $0<x<x_c(q)$, where 
$x_c(q)=1/q$ is the critical curve. \\
\textit{ii)} Rectangles cannot have a scaling function $\mathcal{F}(s)$ 
as in Definition \ref{cr:def:sf} with $s_-\le0$, since the area generating 
function diverges with a logarithmic singularity. This is reflected in the 
above approximation.\\
\textit{iii)} It has not been studied whether the area moments
or the perimeter moments at criticality can be extracted from the above 
approximation.\\
\textit{iv)} The relation of the above approximation to
the area amplitude series of rectangles of the
previous section, $F(s)=\Ei(s^2)\e^{s^2}$, is not
understood. Interestingly, the expansion of $F(s)$
about $s=0$ resembles a logarithmic divergence.
It is not clear whether its expansion at the
origin is related to the asymptotic behaviour of the 
perimeter moment generating functions. 

\subsection{Ferrers diagrams}\index{Ferrers diagrams}

The singularity diagram of Ferrers diagrams is special, since the value 
$x_c(1):=\lim_{q\nearrow1}x_c(q)$ does not coincide with the radius of 
convergence $x_c$ of the half-perimeter generating function $P(x,1)$.
(The function $q\mapsto x_c(q)$ is continuous on $(0,1]$, as may be
inferred from the exact solution.)
Thus, there are two special points in the singularity diagram, namely 
$(x,q)=(x_c,1)$ and $(x,q)=(x_c(1),1)$. Scaling behaviour 
about the latter point has apparently not been studied, see also \cite{cr:PO95}.

About the former point $(x,q)=(x_c,1)$, scaling behaviour is expected.
The area amplitude series $F(s)$ of Ferrers diagrams is given by
the \emph{entire} function
\begin{displaymath}
F(s)=\sqrt{\frac{\pi}{8}}\erfc\left(\sqrt{2}s\right)\e^{2s^2}.
\end{displaymath}
A numerical analysis indicates that its 
Taylor coefficients about $s=0$ coincide with the perimeter moment 
amplitudes at criticality, which characterise a concentrated distribution. 
There is no singularity of $F(s)$ on the negative real axis at any finite 
value of $s$, in accordance with the fact that the critical line at $q=1$ 
extends above $x=x_c$.

It is not known whether a scaling function exists for Ferrers diagrams, or 
whether it would coincide with the amplitude generating function, see 
also the recent discussion \cite[Sec~2.3]{cr:JM06}. An rigorous study 
may be possible, by first rewriting the half-perimeter and area
generating function as a contour integral.  A further analysis 
then reveals a saddle point coalescing with the integration boundary at criticality. 
For such phenomena, uniform asymptotic expansions can 
be obtained by Bleistein's method \cite[Ch~9.9]{cr:O74}.
The approach proposed above is similar to that for the staircase model 
\cite{cr:P95} in the following subsection. 

\subsection{Staircase polygons}\index{staircase polygons}

For staircase polygons, counted by width, height, and area
with associated variables $x,y,q$, the existence of an
area scaling function has been proved. The derivation starts 
from an exact expression for the
generating function, which has then been written as a complex
contour integral. About the point $(x,q)=(x_c,1)$, this
led to a saddle-point evaluation with the effect of
two coalescing saddles.

\begin{fact}[{cf \cite[Thm~5.3]{cr:P95}}]\label{cr:fact:stairsf}
Consider $0<x,y,q<1$ such that the generating function 
$P(x,y,q)$ of staircase polygons, counted by width, height and area, 
is convergent. Set $q=\e^{-\epsilon}$ for $\epsilon>0$. Then,
as $\epsilon\searrow 0$, we have
\begin{displaymath}
\begin{split}
P(x,y,q) &= \left(\frac{1-x-y}{2} +\right.\\
+&\alpha^{-1/2}\epsilon^{1/3}
\left.\frac{\Ai'(\alpha\epsilon^{-2/3})}{\Ai(\alpha\epsilon^{-2/3})}
\sqrt{\left(\frac{1-x-y}{2}\right)^2-xy}\right)
(1+{\cal O}(\epsilon))
\end{split}
\end{displaymath}
uniformly in $x,y$, where $\alpha=\alpha(x,y)$ satisfies the
implicit equation
\begin{displaymath}
\frac{4}{3}\alpha^{3/2}=\log(x)\frac{\log(z_m-\sqrt{d})}{\log(z_m+\sqrt{d})}
+2\Li_2(z_m-\sqrt{d})-2\Li_2(z_m+\sqrt{d}),
\end{displaymath}
where $z_m=(1+y-x)/2$ and $d=z_m^2-y$, and $\Li_2(t)=-\int_0^t\frac{\log(1-u)}{u}{\rm d}u$
is the Euler dilogarithm.
\qed
\end{fact}

\noindent {\bf  Remarks.} \textit{i)} The characterisation of $\alpha^{3/2}$ 
given in \cite[Eq~(4.21)]{cr:P95} has been used.\\
\textit{ii)} The above approximation defines an area scaling function. 
For $x=y$ and $x_c=1/4$, we obtain the approximation \cite[Eq~(1.14)]{cr:P95}
\begin{displaymath}
P(x,q)\sim \frac{1}{4} + 4^{-2/3}\epsilon^{1/3}
\frac{\Ai'(4^{4/3}(1/4-x)\epsilon^{-2/3})}
{\Ai(4^{4/3}(1/4-x)\epsilon^{-2/3})}
\end{displaymath}
as $(x,q)\to (x_c,1)$ within the region of 
convergence of $P(x,q)$. It follows by comparison that the 
area amplitude series coincides with the area scaling function.\\
\textit{iii)} An area amplitude series for the anisotropic
model has been given in \cite{cr:K02}, by a suitable refinement
of the method of dominant balance.\\
\textit{iv)} It is expected that the perimeter law at $x=x_c$ may be 
inferred from the Taylor expansion of the scaling function 
${\cal F}(s)$ at $s=0$. A closed form for the moment generating 
function or the probability density has not been given. 
The right tail of the distribution has been analysed via 
the asymptotic behaviour of the moments \cite{cr:K03,cr:K04}.
See also the next subsection.\\
\textit{v)} The above expression gives the singular behaviour of $P(x,q)$
as $q$ approaches unity, \emph{uniformly} in $x,y$. Restricting to $x=y$, 
it describes the singular behaviour along the line $q=1$ for 
$0<x<x_c$. In the compact percolation picture, this line describes 
compact percolation below criticality. Perimeter limit laws away from
criticality may be inferred along the above lines. (Asymptotic expansions
which are uniform in an additional parameter appear also for solutions
of differential equations near singular points \cite{cr:O74}.)\\
\textit{vi)} By analytic continuation, it follows that the critical curve $x_c(q)$ 
for $P(x,x,q)$ coincides near $q=1$ with the upper boundary curve 
$x_0(q)=(1-s_{-}(1-q)^{2/3})/4$ of the scaling domain, where the value $s_{-}$ 
is determined by the singularity of smallest modulus of the scaling function 
on the negative real axis, hence by the first zero of the Airy function. This 
leads to a simple pole singularity in the generating function, 
which describes the branched polymer phase close to $q=1$. 

\subsection{Self-avoiding polygons}\index{self-avoiding polygons}

In the previous section, a conjecture for the limit distribution of area for 
self-avoiding polygons and rooted slef-avoiding polygons was stated. We 
further explain the underlying numerical analysis, following 
\cite{cr:RGJ01,cr:RJG03,cr:RJG04}. The numerically established form 
Eq.~(\ref{Ecr:ampAk}) implies for the area moment generating 
functions for $k\ne1$ singular behaviour of the form
\begin{equation*}
g^{(sing)}_k(x)\sim \frac{f_k}{(1-x/x_c)^{\gamma_k}}
\qquad (x\nearrow x_c),
\end{equation*}
with critical point $x_c=0.14368062927(2)$ and
$\gamma_k=3k/2-3/2$, where the numbers $f_k$ are related to the
amplitudes $A_k$ in Eq.~(\ref{Ecr:ampAk}) by
\begin{equation*}
A_k = \frac{f_k}{\Gamma(\gamma_k)}.
\end{equation*}
For $k=1$, we have $\gamma_k=0$, and a logarithmic singularity is expected,
$g_1(x)\sim f_1 \log(1-x/x_c)$, with $f_1=A_1$.
Similar to Conjecture \ref{Ccr:SAPlim}, this leads to a corresponding 
conjecture for the area amplitude series of self-avoiding polygons.
If the area amplitude series was a scaling function, we would
expect that it also describes the limit law of perimeter
at criticality $x=x_c$, via its expansion about the origin. (Interestingly,
these moments are related to the moments of the Airy distribution
of negative order, see \cite{cr:RJG04,cr:FL01}.)
This prediction was confirmed in \cite{cr:RJG04}, up to numerical accuracy, 
for the first ten perimeter moments. Also, the crossover behaviour 
to the branched polymer phase has been found to
be consistent with the corresponding scaling function prediction. 
As was argued in the previous subsection, the critical curve $x_c(q)$
close to unity should coincide with the upper boundary curve 
$x_0(q)=x_c(1-s_{-}(1-q)^{2/3})$, where the point $s_{-}$ is related 
to the first zero of the Airy function on the negative real axis, 
$s_{-}=-0.2608637(5)$. The latter two observations support the 
following conjecture.

\begin{conj}[{\cite{cr:Rich02,cr:RJG04}}]
Let $p_{m,n}$ denote the number of self-avoiding polygons of 
half-perimeter $m$ and area $n$, with generating function 
$P(x,q)=\sum_{m,n}p_{m,n}x^mq^n$. Let $x_c=0.14368062927(2)$ 
be the radius of convergence of the half-perimeter generating 
function $P(x,1)$. Assume that
\begin{displaymath}
\sum_n p_{m,n}\sim A_0 x_c^{-m}m^{-5/2} \qquad (m\to\infty),
\end{displaymath}
where $A_0$ is estimated by $A_0=0.09940174(4)$. Let the number 
$s_{-}$ be such that $\left(4A_0\right)^\frac{2}{3} \pi s_{-}$ 
coincides with the zero of the Airy function on the negative real axis
of smallest modulus. We have $s_{-}=-0.2608637(5)$.

\begin{itemize}

\item[\textit{i)}]
For \emph{rooted} self-avoiding polygons with half-perimeter and area 
generating function $P^{(r)}(x,q)=x\frac{{\rm d}}{{\rm d}x}
P(x,q)$, the conjectured form of a scaling function 
${\cal F}^{(r)}(s):(s_{-},\infty)\to\mathbb R$ as in Definition 
\ref{cr:def:sf} is
\begin{equation*}
{\cal F}^{(r)}(s) = \frac{x_c}{2\pi}\frac{\rm d}{{\rm d}s}\log\Ai \left(
\left( 4A_0\right)^\frac{2}{3} \pi s \right),
\end{equation*}
with critical exponents $\theta=1/3$ and $\phi=2/3$. 

\item[\textit{ii)}]
The conjectured form of a scaling function ${\cal F}(s):(s_{-},\infty)\to\mathbb R$ 
for self-avoiding polygons is obtained by integration,
\begin{equation}\label{Ecr:sapsf}
{\cal F}(s) = -\frac{1}{2\pi} \log\Ai \left( \left( 4A_0\right)^\frac{2}{3} 
\pi s \right)+\frac{1}{12 \pi}(1 - q)\log(1 - q),
\end{equation}
with critical exponents $\theta=1$ and $\phi=2/3$.

\end{itemize}
\end{conj}

\noindent \textbf{Remarks.} \textit{i)} The above conjecture is essentially based on 
the conjecture of the previous section that both staircase polygons and
rooted self-avoiding polygons have, up to normalisation constants, the same limiting 
distribution of area in the uniform ensemble $q=1$. For a numerical 
investigation of the implications of the scaling function conjecture, 
see the preceding discussion.\\
\textit{ii)} A field-theoretical justification of the
above conjecture has been proposed \cite{cr:C01}.
Also, the values of $A_1=1/(8\pi)$ and the prefactor $1/(12\pi)$ in 
Eq.~(\ref{Ecr:sapsf}) have been predicted using field-theoretic methods
\cite{cr:C94}, see also the discussion in \cite{cr:RJG04}.

\subsection{Models in higher dimensions}

Only very few models of vesicles have been studied in three 
dimensions. For the simple model of cubes, the scaling
behaviour in the perimeter-area ensemble is the same as 
for squares \cite[Thm~2.4]{cr:J04}. The scaling form in the 
area-volume ensemble has been given \cite[Thm~2.8]{cr:J04}.
The asymptotic behaviour of rectangular box vesicles
has been studied to some extent \cite{cr:MJ05}. Explicit
expressions for scaling functions have not been derived.

\subsection{Open questions}

The mathematical problem of this section concerns
the local behaviour of multivariate generating functions
about non-isolated singularities. If such behaviour is known,
it may, under appropriate conditions,  be used to 
infer asymptotic properties such as limit distributions.
Along lines of the same singular behaviour in the 
singularity diagram, expressions uniform in the parameters 
are expected. This may lead to Gaussian limit laws \cite{cr:FS06}. 
Parts of the theory of such asymptotic expansions have been developed 
using methods of several complex variables \cite{cr:PW06}.
The case of several coalescing lines of different singularities
is more difficult. Non-Gaussian limit laws are expected,
and this case is subject to recent mathematical research \cite{cr:PW06}. 

Our approach is motivated by certain models of 
statistical physics. It relies on the observation that the 
singular behaviour of their generating function is
described by a scaling function. There are major open 
questions concerning scaling functions. On a conceptual 
level, the transfer problem \cite{cr:FO90} should 
be studied in more detail, i.e., conditions under which the 
existence of a scaling function implies the existence of the 
finite-size scaling function.
Also, conditions have to be derived such that 
limit laws can be extracted from scaling functions.
This is related to the question when can
an asymptotic relation be differentiated. Real analytic methods,
in conjunction with monotonicity properties of the generating function,
might prove useful \cite{cr:O74}.

For particular examples, such as models satisfying a linear 
$q$-difference equation or directed convex polygons, 
scaling functions may be extracted explicitly. It would be 
interesting to prove scaling behaviour for classes of 
polygon models from their defining functional equation. 
Furthermore, the staircase polygon result indicates
that some generating functions may have in fact
asymptotic expansions for $q\nearrow 1$, which are
valid uniformly in the perimeter variable (i.e., not only
in the limit $x\nearrow x_c$). Such expansions would
yield scaling functions and correction-to-scaling
functions, thereby extending the formal results of the previous
section. This might be worked out for specific models, at least in the 
relevant example of staircase polygons.

\section*{Acknowledgements}

The author would like to thank Tony Guttmann and Iwan Jensen
for comments on the manuscript, and Nadine Eisner, Thomas Prellberg 
and Uwe Schwerdtfeger for helpful discussions. Financial
support by the German Research Council (Deutsche Forschungsgemeinschaft)
within the CRC701 is gratefully acknowledged.

%%%%%%%%%%%%%%%%%%%%%%%%%
% BibTeX
\bibliographystyle{plain}
\bibliography{refs16}
%
% Non-BibTeX
%\input{referenc}
%%%%%%%%%%%%%%%%%%%%%%%%%

\end{document}